# 湯川秀樹先生の胸像除幕高知訪問3日間の苦悩と生涯の決断：

## 洞窟の蝙蝠から世界へ

Professor Hideki Yukawa's Anguish and a Lifelong Decision During a Three-Day Visit to Kochi to Unveil His First Bronze Statue: From a Cave Bat to the World

大阪大学核物理研究センター

大久保茂男[1]

Research Center for Nuclear Physics, Osaka University, Ibaraki 567-0047, Japan

Shigeo Ohkubo

**要約**

湯川秀樹先生は、高知県の夜須町夜須小学校にPTAが自発的に日本ではじめて建立した湯川秀樹銅像の除幕式に出席するため、5年間の米国での研究生活から1953年に帰国後翌年の1954年3月21日、高知を訪問した。歴史の偶然か、出発の3週間前、3月1日、米国による水爆実験が太平洋ビキニ環礁で行われた。そこでは日本の漁船がたくさん操業していて、事前に知らされることはなく、第五福竜丸をはじめ多くの漁船が被爆した。高知駅に3月21日夕刻到着した湯川は、ビキニ水爆について記者団の質問攻めにあった。湯川は深い苦悩にあった。日本人として「原子の物理」でノーベル賞を受賞した湯川である。湯川は、それは「研究外だ」と質問に答えることを断固拒否した。翌3月22日夕の高知市での一般市民向け講演会では、自分は原子力の研究では素人で専門家はほかにたくさんいる、としてビキニ水爆問題・原子力について発言をすることを前日同様に拒んだ。湯川は、だが、京都に帰り4日後3月28日には有名な「原子力と人類の転機」を起草し、3月30日新聞発表した。以後、ビキニ水爆問題・原子力問題の激流に引き込まれていく。苦悩の湯川はいつ決断したのか。本稿では何が懊悩の湯川をして短期間、1日のうちに決断させるにいたったか、なぜ一夜で大きな心境の変化が起こったのか、資料に基づき詳細に明らかにされる。

目次

---

[1] Research fellow of RCNP, Emeritus Prof. of The Univ. of Kochi, 高知県立大学名誉教授

## 第 1 章　はじめに

　拙稿[1]で湯川秀樹先生(1907-1981)のノーベル賞受賞を記念する日本で初めての湯川胸像が高知県の小学校に住民の自主的運動で建設され、湯川が夫人とともに除幕式に出席するため 1954 年 3 月に高知を訪問していたことを報告した。アメリカでの 5 年間にわたる研究生活を終え前年夏に帰国して間もなく、多忙な中での出席であった。四国で唯一訪れていない高知をぜひ訪問したい、また恩師で元三高校長の森総之助(1876-1953)の出身地土佐への訪問願望もあって実現した。

夫人の強い希望もあった。湯川には楽しい「四国の春」の旅となるはずであった。

　私はこの稿で、世に知られていない湯川銅像（浜口青果（1895-1979）作）の存在と湯川夫妻の訪問のみを記し、それ以外の事柄には意図的に触れなかった。執筆のため資料を調査する中で、この旅が、湯川にとって人生の決断を迫られる苦悩の旅であったことを知った。湯川に憧れ、京都大学物理学科に進み、湯川の教えを受け理論物理学・原子核理論研究の道に進んだ私にとって、40歳代の湯川が高知訪問で大変苦悩していることを知ったことは大変な衝撃であった。世の中では、湯川は日本人として初めてノーベル賞を受賞し、敗戦で塗炭の苦しみの中にあった国民に勇気と希望と自信を与え、国民に敬愛され、多くの人々が思っていたように順風満帆の人生をおくったものと思っていた。1954年ごろから関わる核廃絶の運動も科学の平和利用などの高尚な理念を実現するための科学者としての自覚的な活動だとばかり思っていた。

　ところが、事実はそうではなく、湯川は私の故郷である高知訪問中に大変な苦悩のなか、清水の舞台から飛び降りるほどの一大決心をして、核廃絶・平和のための運動の道に進んだのであった。世の中には、湯川にこのような深い苦悩があったことを示す資料は何ひとつなく、誰も知らなかった。私はこの湯川の苦悩については、自分の中にのみ留めておこうと思い、拙稿[1]では、湯川の楽しい高知訪問のみを記し、その稿の最後を次のように締めくくったのも、そうした理由からであった。「湯川秀樹先生の胸像建立と除幕式は町立夜須小学校6年生の卒業式にあわせ計画されたという。湯川秀樹はその日、第三高等学校時代に物理の手ほどきを受けた恩師・森総之助の生地、高知のまちを訪ねた。1954年3月22日、高知県夜須町の春は子どもたち・町民たち・除幕式出席の湯川秀樹夫妻にとって日ざしのやわらかい春分の日の、俸あふれしあたたかい『高知の春』、『四国の春』そして『日本の春』であった。湯川の随筆『四国の春』は残されず、湯川胸像はその後語られることがなかった」[1]

　私は拙稿[1]を発表以降、本来の理論物理学の研究に戻り、折しも2020年から突然始まったコロナ禍で、他の人々と同じく、政府や自治体による「不要不急の外出自粛」要請とその社会的雰囲気の中で、社会から隔絶を強いられ家にこもる生活となった。2019年5月に『素粒子論研究電子版』に公表された拙稿[1]は、「オープンアクセス」で公開されていて素粒子・原子核研究者をはじめ、関心のあるひとはだれでも自由に読むことができる。科学史などの専門学術誌ではなく、『素粒子論研究電子版』を選んだのも、素粒子・原子核を中心に物理学研究者や関心のある方に自由に読んでもらいたいと思ったからであった。

　論考[1]の公表から2か月後、新聞記者（朝日新聞・湯川うらら）から、湯川胸像除幕式の写真が新たに見つかったと連絡があり、新聞社に赴くと、湯川が胸像除幕式に臨み、自身の胸像を見上げている鮮明な、これまで知られていなかった写真3枚を見せてくれた。若い新聞記者は私の論考[1]を読んですぐ、2019年初夏に現地夜須町を訪れ、持ち前の行動力で、当時の夜須小学校6年生で写真の所持者を見つけたのである。私が知らなかった湯川胸像が除幕される瞬間の写真が卒業生宅に保存されていたのである。2019年夏のことである。うら若い記者に、「いい原稿を書いてください」と伝え、1時間ほど取材に応じ、新聞社を後にし、夏の日差しが強い中を家路についた。その後、このことはすっかり忘れて、理論物理の研究に戻って没頭していた。

コロナ・パンデミック下で家にこもる状態が続いていた2021年1月、湯川胸像除幕式に出ていて湯川の話を聞いた、当時夜須小学校6年生であったという清藤禮次郎から突然の電話がかかってきた。湯川胸像について資料を持って説明したいとのことであった。私は湯川胸像除幕式を歴史的事実としてすでに過去のことだと考えていたので、当時の関係者が現実空間に突然現れたことに驚愕した。2021年2月のある寒い早朝、コロナ禍の中、拙宅まで来られ、たくさんの資料を見せてくれた。その後、再び寒い早朝に、除幕式で児童代表として除幕を務めた浜田英子（旧姓春樹）を伴って訪ねて来た。当時のことを思い出し、持参したアルバムを出して、湯川胸像除幕式のことを話してくれた。

　私は一年ほど考えて、新たに見つかった湯川夫妻、そして湯川胸像の写真を含めて、第2稿を書くことにした[2]。卒業生、夜須小学校関係者、新聞記者の熱意におされ書かずにはいられない思いであった。2022年3月公表のこの稿[2]で、私は初めて湯川の「初めての高知への旅」が、喜ばしい記念の「胸像除幕式」にとどまらず、湯川の生き方をも変える苦悩と生涯の転機の旅であることに触れた。高知訪問は、湯川の社会における立ち位置が大きく変わる後半生の出発となる旅であった。この「湯川秀樹先生の生涯の転機」に関する論考[2]は新聞でも報道され[3] [4]、折しもロシアがウクライナ侵攻で核兵器による威嚇を行っている状況と重なり、社会の関心を呼んだ。湯川の研究一筋の人生からの大きな苦悩の決断を、世の人々が初めて知ることとなった。「湯川秀樹」を研究している専門家によっても、驚きをもって注目されることとなった[5] [6]。

　この第2稿[2]でも私は、なぜ湯川が大きな決断を短期間のうちに行うことになったのか触れなかった。湯川には大きな葛藤があり、第三者が外部から安易に論じたり、決めつけたりできるものではないので、これ以上言及するのは、控えたいという思いがあったため、再び理論物理の「原子核虹」と「原子核のクラスター構造」の研究に没入していた。

　人の人生には、一日にして大きな決断を迫られることが、生涯のうちに何度かある。「花木春過ぎ夏すでに中ばなり」（細川頼之(1329-1392)「海南行」）というが、筆者にはもう「秋半ばなり」である。28万人以上もの人々に閲覧され、社会的な反響も大きいこともあって、湯川先生が苦悩のなか、どうして短期間のうちに研究者としての生涯を変えるような大きな決断をしたのか触れておき、後世のために書き残しておきたいと思うようになった。「湯川秀樹はノーベル賞を受賞したあと、早々に研究の現場から退き、活動の場を核兵器廃絶運動など学術の外に求めた」[7]といった誤解があることなどを鑑みると、湯川が戦後の時代背景のなかで苦悩の末に決断することになった経緯を記すことは、いっそう重要であると思うようになった。湯川の教えを受けた者として、また高知に生まれ湯川の苦悩を知った者として、これは他に代えがたい私の務めとして記そうと2018年以来7年の苦悩の末に「決意」した次第である。湯川はなぜ、学問一筋の研究生活から社会的な関わりを強いられ、その決断を下したのか。私は明治生まれの人の「時代を担う気骨」に打たれた。高知訪問当時の時代背景を思い起こしつつ、その経緯を探ってみたいと思う。本稿が前2稿[1] [2]と独立に理解できるよう、要点を振り返りながら稿を進めていきたい。

## 第2章　高知訪問前の湯川秀樹：中間子論研究と米国からの帰国

湯川の苦悩と決断を理解するには、彼が生きた社会・時代的背景の理解なしには不可能であろう。湯川の物理学研究、ノーベル賞受賞、そして当時の社会的状況を思い出しておきたい。

　まず、簡単に中間子論研究、ノーベル物理学賞受賞、そして敗戦からの戦後復興を概観する。湯川秀樹は、イギリスの物理学者チャドウィック（J. Chadwick、1891-1974）による1932年の中性子発見以来、原子物理学における大きな問題であった原子核が安定に存在する理由を理解するため、陽子や中性子間に働く核力が未知の新粒子である「中間子」によって引き起こされているとする中間子論を、1935年に日本の学会誌に発表した。予言された粒子は宇宙線中に探索され、1938年には湯川粒子、ユーコンなどと呼ばれ、湯川は世界的に注目されるようになった。いろいろな呼び名がつけられた中で、旧制高知高校（現高知大学）教授の篠崎長之（寺田寅彦（1878-1935）の弟子）が1939年に雑誌「科学」[8]で提唱した「中間子」の名称が定着している。湯川の予言した核力を媒介する中間子は、1947年にイギリスの物理学者パウエル（C. F. Powel、1903-1969）らによって宇宙線中に発見され、中間子論の正しさが明確になった。素粒子物理学を切り拓き、1949（昭和24）年、日本人として初めてノーベル賞を受賞し、敗戦で打ちひしがれ、科学者の中でも「『虚脱』という言葉が氾濫し＜略＞未来に希望をもてるような意見はまったく聞けず、社会全体に不安と空腹のみが満ちていた」[9]時代の国民に勇気と自信と希望を与えた。

　湯川は戦後、理論物理学者としてだけでなく、核兵器廃絶・平和運動のために後半生を捧げた[2]。湯川が物理学者として原水爆、核兵器、平和問題について社会的に意見を初めて表明したのは、1954（昭和29）年3月30日の毎日新聞においてである[10]。

　米国に5年間滞在していた湯川は、1953年7月に帰国する。ノーベル賞受賞を記念して京都大学に設立された湯川記念館に、全国初の共同利用研究所である基礎物理学研究所が設立され、その所長となるためである。帰国直後の9月には、基礎物理学研究所などを会場に理論物理学国際会議が開催され、著名な物理学者（ノーベル賞受賞者17人（後の受賞者を含む））が多数参加した。これは戦後日本の国際社会復帰への大きな一歩となる。湯川は国際会議の会長を務めた。米国の原爆開発のマンハッタン計画を主導し「原爆の父」と呼ばれる共同会長のオッペンハイマー（J. R. Oppenheimer、1904-1967）は直前になって来日を中止し、代わりに素粒子物理学者のパイス（A. Pais、1918-2000）が来た。彼は国際会議参加時の印象記の中で、「湯川は日本で天皇に次いで有名な人物だったことに注意していただきたい」[11]と記した。敗戦後の国民にとって、湯川は救世主のような存在であっただろう。2024年の最近の新聞投書でも、「敗戦後のみじめな時期に希望の光となった博士」[12]と記されている。社会と政治的に関わることを意識的に避け続けてきた湯川にとって、歴史の展開は容赦なかった。初めての高知訪問がその「序幕」であった。高知県夜須町の「湯川胸像の除幕」は湯川の人生を大きく変える、新たな困難な「除幕」でもあった。湯川に関する伝記・研究書は数多あるが、筆者の論考で明らかになるまで湯川の苦悩の人生は気づかれなかった。

## 第3章　湯川秀樹の高知訪問：受諾から高知訪問まで直前の状況

　湯川胸像の建設については高知新聞1953年12月6日[13]に「夜須町小学校PTA委員会が湯

川秀樹博士胸像について協議」とあり、高知新聞1953年12月28日[14]には「湯川博士の胸像建設へ」と報じられている。湯川が高知へ来ることは1954年1月12日までには親友の大岡義秋の尽力で決まり、高知新聞は1954年1月13日付で次のように報じている[15]。「湯川博士三月中に来高　高知市中央公民館では科学教育振興のため湯川秀樹博士招聘を計画、京都大学医学部会に出席した同博士の中学・高校時代のクラスメート大岡義秋高知赤十字病院長にあっせんを依頼したところ十二日帰高した大岡院長から三月二十日ごろ博士夫妻そろって来高を快諾したとの連絡があった。片岡公民館長談　「さっそく博士を本格的に迎える計画をたてている、第一日は教育者のための原子力の講演、第二日は一般市民の通俗講演を予定しており、婦人の部では湯川夫人を囲んで懇談会を催したい」（下線は筆者）

　ここで注目すべきは、湯川は大岡の要請を「快諾」したという点である。日程は確定していないものの、一度で快諾が得られたのは大岡と湯川が親密であったことに拠っている。我が国初めてのノーベル賞受賞者であり、帰朝後に全国から講演依頼が殺到したであろうことは容易に想像される。その後日本ではノーベル賞受賞者は何人かでるが、「講演は原則断るのが肝要だ」と、友人の益川敏英（2008年ノーベル物理学賞受賞者）から聞いたことがある。ましてや、帰国からまもなくの、初めての遠隔地、高知訪問である。この大岡の存在が、湯川の世紀の「決断」に影響を与えた可能性は、後に議論する。高知新聞1954年2月6日[16]は、「夜須小に湯川博士胸像卒業式に除幕、制作は浜口青果氏」と次のように報じている。「かねてから科学教育の振興を熱心に唱えていた香美郡夜須小学校PTA川村会長は子供たちがあやかるようにノーベル賞を受けた湯川博士の胸像をたてることを思い立ち昨年同地に一時居住していた彫刻家浜口重蔵青果さんに制作を依頼、浜口さんも"そういうわけなら"と石膏像が出来上がり、三月早々に胸像が完成、同月二十四日の卒業式当日に除幕式が行われことになった。なお、関係者は除幕式には講演のため来高交渉中の湯川博士の出席を希望している」。この記事から、湯川は高知訪問を快諾したものの、湯川胸像除幕式への出席はこの時点では決まっていないということが分かる。

　最終的に高知訪問が確定するのは2月19日[17]で、高知新聞は次のように報じている。「湯川博士の来高本決り　高知市では日本の生んだノーベル賞受賞者湯川秀樹理博（京都大学理学部およびコロンビア大学教授）の講演を要請していたが、このほど同博士から快諾の返事があった、県、県教委、市、市教委および高知新聞社共催の講演会は三月二十二、三両日市中央公民館で開かれるが、同博士の滞在中の日程は次のように決った（ママ）。二十二日＝午前十時半－同十二時、香美郡夜須町小学校の湯川博士胸像除幕式に参列ののち同郡赤岡町城山高校で講演、午後六時半－同八時半、一般講演（市中央公民館）▽二十三日＝午前十時－同十一時市内小、中学生を対象とした講演会（交渉中、市中央公民館）午後二時－同四時学術講演会（市中央公民館）」

　この時点で、夜須小学校の除幕式出席も決まった。大岡が湯川との交渉だけでなく、講演会での企画に関しても重要な役割を果たしていることは、次の学童講演会についての高知新聞1954年2月24日[18]の報道でも分かる。「湯川博士に激励頼む　三月に来高する湯川秀樹博士に昨年十一月行った子供科学展に特選で入選した小、中学生の作品をみてもらい小さい科学者たちを激励してもらうことになった。同博士の中学、高校時代のクラスメート大岡高知赤十字病院長の

話によると大正十一年冬、世界的学者アインシュタイン博士が来日したとき中学生であった湯川少年がその講演を聞き深い感銘を受け、この時から物理の世界に志をたてたということで大岡院長と片岡中央公民館長の肝いりで子供科学展特選組二十名を選んで来月の二十三日午前十時に行われる小、中学生に対する講演会終了後博士を囲んで懇談、博士のあとにつづく少年を本県からもだそうというわけ」

　こうして、湯川の高知訪問の計画は着々と進み、1954年2月28日（日曜日）の高知新聞一面中央に四段記事で、湯川訪問と講演会が主催・共催者連名で次のように大きく報じられた（図1）[19]

**湯川秀樹博士の講演会**
**三月二十二、二十三日、市中央公民館**

一九四九年度ノーベル物理学賞に輝く湯川秀樹博士夫妻を招き講演会を次の要領により開催いたします。<u>原爆、水爆あるいは原子力の管理とその平和利用が世界の論議の焦点となっている今日</u>、まことに意義あるものと確信いたします。また当日は非常な盛会が予想されますが、公民館の収容力に限度がありますので聴講者は場外にあふれることも考えられ特に場外マイクの設備をいたしますので多数の来聴を期待します。

【会場】高知市中央公民館
【日時】①三月二十二日「一般講演」、午後六時半から八時半まで
　　　　②二十三日「学術講演」午後二時から四時半まで
【聴講入場券】　一般講演＝▽学生（高校、大学）は県教委指導課へ申込み人員割り当を受けて下さい　▽一般＝市公民課、市公民館、高知新聞社で聴講券を渡します。（共に収容人員に制限がありますので聴講券を入手できない場合は、悪しからずご了承下さい。
学術講演＝大学教授、大学生、高校教諭（共に物理学専攻）大学関係は高知大学へ高校は県教委指導課へ申込むこと（人員予定　五十名）
【小、中学生のための講話】▽二十三日午前十時＝入場券は、県教委、市公民課で各学校人員割当を受けること
【夫人を囲む座談会】令息病気のため出席は未定です。開催不能の場合はご了承ください。
　なお博士は二十二日午前十一時半、夜須小学校の博士胸像除幕式に参列することになっています。
　　　共催　高知県、高知市、県、市教育委員会、高知新聞社

</div>

（下線は筆者）

　翌日、週明けの月曜日、3月1日にアメリカによる水爆実験が行われることになろうとは、この講演会案内が出た時点では湯川をはじめ誰も夢想だにし得ないことであった。歴史には何故偶然が重なるのか。ビキニ環礁水爆実験と時を同じくして、1954年3月2日には原子炉築造予算案、

すなわち『原子炉築造のための基礎研究および調査費』が突如国会に提出された。改進党の中曽根康弘(1918-2019)代議士らによるウラン 235 にちなんだ 2 億 3500 万円の原子炉築造予算は、国内の出来事として新聞報道され、湯川を含め国民が知ることのできる事象であった。

　この「湯川博士の講演会案内」に「原爆、水爆あるいは原子力の管理とその平和利用が世界の論議の焦点となっている今日、まことに意義あるものと確信いたします」(下線は筆者)とあるのは、中曽根代議士らによる原子炉築造予算の成立や、ビキニでの水爆実験による第五福竜丸の被爆が広く知られる前の 1954 年 2 月 28 日の時点においても、湯川に対する原爆・水爆についての見解を講演会で知りたいという機運が国民の中に強かったことを示している。3 月 1 日のアメリカによるビキニ環礁での水爆実験自体は、第五福竜丸の被爆が読売新聞[20]でスクープ報道されるまでは国民的な関心事とはならず、湯川の高知訪問計画は 3 月に入ってからも順調に進んでいった。

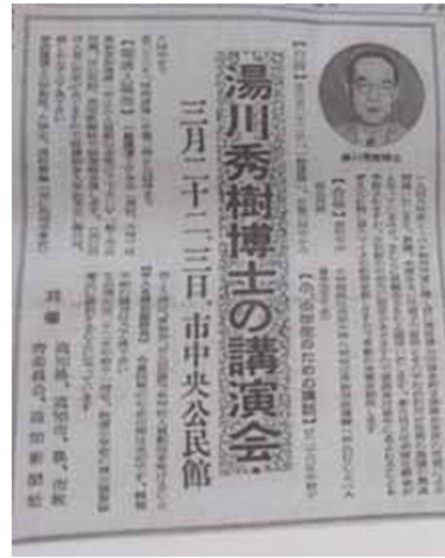

図1　湯川秀樹博士の講演会の新聞案内、高知新聞１９５４年２月２８日[19]

　3 月 3 日、高知新聞は湯川の夜須小学校への除幕式出席日程の詳細を「町から村から」欄[21]で次のように報じている。「夜須小学校玄関に建設の湯川秀樹博士胸像除幕式は二十二日午前十時三十分博士を迎えて行う予定になっている」

　さらに、高知新聞 1954 年 3 月 6 日[22]は湯川歓迎の三高同窓会の会合の計画が進められており、大岡義秋がかかわっていることを次のように報じている「県、市の歓迎計画はじめ在高の三高会では平田病院長、宮崎県農林部長、大岡高知赤十字病院長、浜田市建設部長、五十嵐高知大教授、二宮高知労働基準局長ほか二十名の同窓生が集まり二十三日午後六時から五台山荘で同窓のつどいを催すなど歓迎の方もひっぱりだこ、また去る二月二十八日京都の湯川博士を訪問してこのほど帰高した二宮高知労働基準局長は"たぶん博士夫人も来高するでしょう"と次のように語った。『湯川博士は恩師故森総之助氏＝元三高校長、本県出身＝に物理の手ほどきを受けたので高知は特になつかしい、妻も高知行きを希望しているので出来れば同伴、龍河洞、桂浜などの景勝地を見物したいと来高の日を楽しみにしていた。』」

　こうして、2 月 28 日の講演会の新聞公告の時点では、息子の病気のため未定であった湯川夫人の高知訪問も確定した。湯川夫人(澄子)(1910-2006)[23][24]のこの高知訪問への同伴が、湯川の「生涯の決断」にとってきわめて重要であっただろうことは、後に論じられる。

　除幕式については湯川の出席が 3 月 3 日までには確定したので、準備が進められ、3 月 9 日には湯川博士歓迎と除幕式の準備会が開かれたと、高知新聞 1954 年 3 月 13 日[25]が報じている。「夜須町教育委員会および小学校 PTA 役員は九日午後一時町役場会議室で湯川博士の胸像除幕式と同博士歓迎準備について協議した」

そして、1954年3月12日までに湯川夫妻の高知訪問が確定したことが、高知新聞1954年3月15日[26]に報じられている。「湯川夫人の来高決る　湯川博士夫人澄子さんの来高が決った。これは主催者側の問合せに対して十二日、京都の湯川博士から大岡高知赤十字病院長宛に夫人を同伴すると回答があったもので湯川夫人を囲む座談会は二十三日午後一時半（場所未定）から行われる予定」。湯川の息子（湯川春洋）の病気で夫人の訪問決定はずれ込んでいたが、夫人の強い希望が叶い、実現することとなった。

湯川からの連絡は大岡義秋との間で行われていることが、ここでも分かる。湯川と大岡義秋は京都で中学・高校の親友である。この大岡義秋も湯川の「生涯の決断」において重要な存在であることが、後に論じられる。

夜須小学校での湯川胸像の設置は3月12日に行われたと、高知新聞1954年3月16日[27]が報じている。「夜須小学校玄関側に建設の湯川秀樹博士の胸像は12日同町出身の彫刻家浜口青果氏から送られたので工事に着手した」。こうして湯川の高知訪問の計画は、大岡義秋を窓口に順調に進められ、湯川の3月21日の訪問を待つばかりであった。

## 第4章　高知訪問決定後の原子力・原爆をめぐる社会状況の激変

ビキニ事件[28][29]とは、1954年3月1日にアメリカがビキニ環礁で水爆実験を行い、マーシャル諸島で操業していた日本の遠洋マグロ漁船第五福竜丸（静岡県焼津）の船員23名全員が水爆の放射性降下物で被曝し、無線長の久保山愛吉（1914-1954）さんが放射線障害で死亡したことが、今日でも広く知られている事件である。ビキニ環礁近くでは、多くの日本の漁船が操業していた。遠洋漁業が盛んな高知県からも多数の漁船が進出・操業していた[29]。

ビキニ水爆実験が世に知られるようになったのは、先に述べたように、3月16日に読売新聞が朝刊で「邦人漁夫、ビキニ原爆実験に遭遇　23名が原子病　1名は東大で重症と診断」とスクープ報道してからのことである。

1954年3月2日に突如原子力予算が国会に提出され、日本政府による原子力の推進は、これだけでも湯川が社会から様々な意見を求められる機会を増やすものであった。しかし、ビキニ水爆実験は湯川にとって、その研究人生をも変えうるほどの、更なる大きな衝撃を与えるものであった。

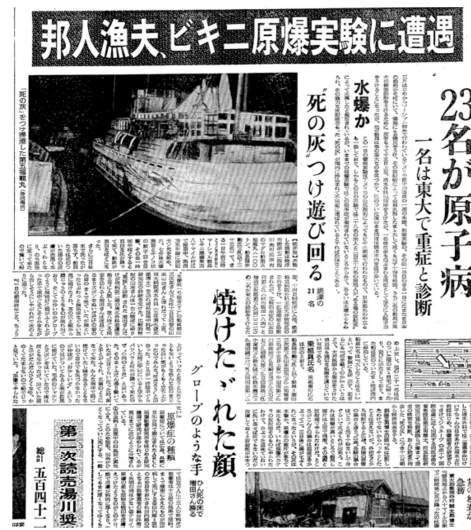

図2　ビキニ水爆で被爆した日本漁船を報ずる読売新聞1954年3月16日朝刊[20]

ビキニ水爆実験は3月1日に行われたが、それがすぐに世の中に衝撃を与えたわけでもない。1954年3月1日からの状況を、やや詳しく見てみる必要がある。最初に新聞報道されたのは3月2日のことである。この時点では「水爆」という言葉は使われていない。朝日新聞の3月2日付

夕刊は一面でこそ報じているが、「新原子爆発実験 マーシャル群島で始まる」と小さく伝えるに留まっている。その後、3月14日付夕刊に「強力水爆を実験 米政府高官ほのめかす」という記事が出ている。3月16日には「ハワイへ出発 原子力委員長」という記事も掲載されている。3月14日に第五福竜丸が焼津港に帰港するまで、大きな騒ぎは全くない。当時、原爆や水爆は「原子力」と呼ばれ、アメリカでも原子力委員会が管轄し、水爆実験も原子力委員会の予算のもとで開発・実験が行われていた。3月1日の最初の発表もアメリカ原子力委員会のストローズ委員長が行っており、「原子力装置」の爆発がマーシャル諸島で行われたと発表している。

3月14日に第五福竜丸が帰国し、3月16日に被爆が明らかになった。読売新聞16日朝刊[20]がスクープで報じた「邦人漁夫、ビキニ原爆実験に遭遇、二十三名が原子病」の記事で世界がビキニ水爆実験の惨事を初めて知ることになる。記事は次のように伝えている。「遭難した漁船は、焼津市焼津七二四西川角市氏所有のマグロ船第五福竜丸（百トン、船長筒井勲氏[二四]以下二十三名）で、さる一日午前三時ごろ（現地時間）マーシャル群島ビキニ環礁東方八十マイル付近で操業中、南西方水平線に突如セン光を認め、六、七分後大爆発音とキノコ状の原子雲を目撃、その後一時間半ほどすると真白な灰が降って来て船体が真白なほどになった。さらに三日後水で洗ったが落ちず、船員たちは灰のついた部分の皮膚が赤黒く水ぶくれとなり、のちに黒色に変わって来たので驚いて帰国、一四日朝焼津に入港、焼津協立病院外科主任大井俊亮氏の手当てをうけ、一応原爆症と診断されたが、比較的重症の山本忠司君（二六）（同市焼津一八四）と増田三次郎君（二九）（同二二二七）の二名が東大の診断をうけるため上京し、他の二十一名は灰のついた服のまま自宅に帰ったり、遊びに出たりしており、また船は灰のついたまま焼津港内に停泊している。船員たちは『大したことはない』といって警察にも届けず、記者らにも『何を騒ぐのか』というほどで、大井医師から報告をうけた県当局も適切な処置を講じていない有様である」

湯川は3月16日当日すぐに日記に書いている[30]。高知へ出発する1週間前である。

> 「3月16日 火 曇
> ＜略＞三月一日ビキニ環礁北東約百マイルの地点で水爆実験による真っ白な灰を被ったマグロ漁船第五福竜丸帰港、火傷の傷害を受けた乗組員を診断 水爆症と推定」

このことは、湯川がビキニ水爆実験と乗組員の被爆に大きな関心を持っていたことを示している。（湯川の日記に、この件に関する記述は、これ以降、彼が高知から帰洛した5日後の3月28日まで現れない）

表1に、湯川が高知へ出発するまでのビキニ水爆実験に関する新聞報道がまとめられている。湯川が高知を訪問するのは3月21日－23日であり、この間のことは日記にない。湯川に出発直前になって、予想もしない大事件が起きたのである。湯川は3月16日の読売新聞報道以降から出

| | 表1　ビキニ水爆実験をめぐる新聞報道 | | |
|---|---|---|---|
| | 掲載紙 | 掲載日 | 新聞見出し |
| 1 | 朝日新聞 | 1954年3月2日夕刊 | 新原子爆発実験　マーシャル群島で始まる |
| 2 | 朝日新聞 | 1954年3月8日朝刊 | "破壊力時代が到来"　ボールドウィン氏論ず |
| 3 | 朝日新聞 | 1954年3月14日朝刊 | 強力水爆を実験　米政府高官ほのめかす |
| 4 | 朝日新聞 | 1954年3月16日朝刊 | ハワイへ出発　原子力委員長 |
| 5 | 読売新聞 | 1954年3月16日朝刊 | 邦人漁夫、ビキニ原爆実験に遭遇　23名が原子病　1名は東大で重症と診断 |
| 6 | 朝日新聞 | 1954年3月17日朝刊 | ビキニの灰　三度味った原爆の恐怖 |
| 7 | 朝日新聞 | 1954年3月17日夕刊 | "不幸な出来事"　ダレス長官語る |
| 8 | 朝日新聞 | 1954年3月17日夕刊 | 駐日大使館からの報告待つ　米政府スポークスマン談 |
| 9 | 朝日新聞 | 1954年3月17日夕刊 | 警戒措置はしたと思う　ス原子力委員長語る |
| 10 | 朝日新聞 | 1954年3月17日夕刊 | 調査を正式に要求　ビキニ問題　島公使、国務省に |
| 11 | 朝日新聞 | 1954年3月17日夕刊 | 水爆は運べる　米両院原子力委　コ委員長言明 |
| 12 | 朝日新聞 | 1954年3月18日夕刊 | 同海域にいた理由調査　コール委員長談 |
| 13 | 朝日新聞 | 1954年3月18日夕刊 | ビキニ「水爆」実験の真相　想像絶した爆発力　米科学陣も驚倒 |
| 14 | 朝日新聞 | 1954年3月18日夕刊 | 米二、三日中に回答　国務省で表明 |
| 15 | 朝日新聞 | 1954年3月18日夕刊 | 威力公表に両論 "世界平和の教訓に" 両院原子力委員 |
| 16 | 朝日新聞 | 1954年3月19日夕刊 | 米からコウ薬　原爆火傷用に |
| 17 | 朝日新聞 | 1954年3月20日朝刊 | 直ちに責任を調査　両院原子力委コール委員長語る |
| 18 | 朝日新聞 | 1954年3月20日朝刊 | 「ビキニの灰」の正体　東大調査班一端をつかむ　原爆のカケラもまじる |
| 19 | 朝日新聞 | 1954年3月20日夕刊 | 立入禁止海面を拡大　原子力実験で日本政府に通告 |
| 20 | 朝日新聞 | 1954年3月20日夕刊 | 次回実験は事前に警告 |
| 21 | 朝日新聞 | 1954年3月21日朝刊 | 水爆被災・第一号か |

発までのあいだに、京都において新聞社などから取材の申し込みを受けていたことであろう。

　ビキニ水爆事件にもかかわらず湯川の高知訪問の予定は変わらず、粛々と進められた。湯川は大岡義秋に約束した高知訪問の約束を誠実に守った。3月21日の高知新聞[31]は湯川の高知訪問が当初の予定通り行われることで、次のように報道した。「湯川博士夫妻　あす南風で来高　湯川秀樹博士は澄子夫人同伴で21日午後7時15分高知駅着の列車で来高、22日午前10時半夜須小学校の博士胸像除幕式に参列するが帰途とくに午後零時半から城山高校講堂で約1時間講演、同6時から高知市中央公民館の一般に対する講演で第1日の日程を終り、23日は午前11時から1時間学童に対し正午から午後1時20分まで県商工会議所で高知ロータリークラブ会員にそれぞれ講話、午後1時30分から高知中央公民館で2時間にわたって教職員などに最後の学術講演を行う。

博士夫人澄子さんは終始博士と行動を共にするが23日午前11時から2時間、三翠園内水哉閣の婦人団体主催の婦人を囲む会に出席する。なお博士の宿舎は高知市本丁筋の城西館、23日午後3時58分高知発の列車で離高する」

湯川は国鉄の列車で陸路京都から高知へ来た。1954年当時の旅行経路は、京都から東海道線で岡山経由宇野港まで行き、そこで宇高連絡船で高松へ渡る。高松からは土讃線を高知まで行く経路である。筆者の記憶では高松－高知間は蒸気機関車であり、四国山脈を越えるのはトンネルが多く時間もかかる。午後7時15分に高知に着くには京都を朝早く出発しなければならない。当時の国鉄の時刻表[32]で調べてみた。夕方19時15分に高知駅に到着の列車は土讃線高松駅15時20分発 の「2・3等 準急 南風」しかない[31]。この列車に接続する宇高連絡船は宇野港発13時58分発高松桟橋着15時08分着の連絡船[33]である。この連絡船に間に合う列車は東海道本線・山陽本線の京都発8時49分で岡山着12時55分、宇野着13時45分着の急行列車[34]がある。これは東京駅22時30分発宇野行きの夜行列車である。当時は移動に航空機を使うことはない。結局、湯川は京都駅を8時49分の急行列車でたち、19時15分に高知駅に到着したことになる。社会のテンポが緩やかな時代とはいえ一日を移動にかける長旅である。

## 第5章　高知訪問中の湯川発言と時系列変化（3月21日―23日）
### 5.1　3月21日夕方　高知駅到着時の記者会見での発言

長旅で疲れたであろう湯川夫妻は高知駅で高知県民から大歓迎を受けた。高知市長をはじめ1000人を超す市民が出迎え、駅は溢れんばかりであった。湯川にとっては生まれて初めての高知訪問である。翌日3月22日の高知新聞朝刊[35]は湯川の高知駅到着と歓迎の様子を次のように伝えている。「一九四九年度ノーベル物理学賞に輝く湯川秀樹夫妻は高知新聞社主催の講演会ならびに夜須小学校同博士胸像除幕式出席のため二十一日午後七時十五分高知駅着"南風"で来高した。博士はグレイの背広に水玉模様のネクタイ、柔かい黒のオーバーをきこみ顔色もよく、若々しい澄子夫人はコン地の花模様の和服姿で出迎えの氏原市長、高橋県教育長、伊藤市教育長、日赤大岡博士、福田高知新聞社社長、堅田同編集局長、教育関係者、婦人団体、学生ら約千名にニコやかに挨拶した」

出迎え者の中に大岡がいることに留意したい。ここに名前の挙がっている出迎え者の中で、湯川と中学・高校から懇意なのは大岡のみである。高知駅から宿泊の城西館まで親しく案内できるのは大岡しか見当たらない。

「到着後、湯川博士は駅長室で記者団との一問一答の記者会見に臨み、その内容は以下の通りである。
　－　高知の印象は
　　博士　本県は初めてです。四国で高知だけ来ていないので今回実現したのはうれしい。
　－　ビキニ被爆と政府の原子炉予算[2]について

---

[2] ここでの原子力予算は、1954年3月（昭和29年）、改進党の中曽根康弘代議士らが、日本での原子力発電研究のための予算2億3500万円を、アメリカのビキニ水爆実験のあった翌日の3月2日、国会に提出したものを

博士　これについてはご承知のように私は一般に言明したことがなく全く関知しないところで私の研究外だ、この問題は答えられない
　　―　原子力の平和利用について
　　　博士　原子力の国際管理は必要なことで日本も将来参加の要請があれば参与すべきだ、原子力を平和的に使うべきだということはいまさらいうまでもない
　　―　高知での講演内容は
　　　博士　原子力についてはふれない、一般的な所見、科学者としての体験について述べるつもりだ。
　　―　夜須小の胸像について
　　　博士　私は皆がいうほど偉人でもなければりっぱな人間でもない。二宮尊徳像にかわって私の胸像建立はどうかと思うが日本では最初のことだし出席することにした。＜略＞」[35]（下線は筆者）

　会見記事は続く。「ついで澄子夫人は科学者の妻として感想を次のとおり語った。研究に熱中しすぎて家の中が暗くなり易いのでつとめて明るいふんい気を作るように努めています。主人は胃腸が弱いので食物はとく注意し、睡眠不足にならぬよう留意しています。中間子の存在も寝床で発見したくらいなので頭をつかれさせぬよう心がけています。科学者の妻といっても主人に来た手紙を整理するくらいのことで一般とそんなに変りはありません」[35]

　注目されるのは新聞記者のもっぱらの関心が、ビキニでの水爆実験と関連する原子力の問題であることである。湯川は新聞記者の質問には研究外だとして答えていない。妻が記者会見で話しているのも注目される。

　2 日後、1954 年 3 月 23 日（火曜日）高知新聞の夕刊コラム『話題』「湯川さん」（1 面）で松田記者は次のように書いている[36]。「湯川博士が来高した。ノーベル賞の湯川といえばこどもでも知っているせいか、高知駅頭は大した歓迎ぶりだった。ちょうどビキニの第五福竜丸事件で原子力への関心が一段と高まっている折だけに、原子物理学の権威、湯川博士の顔を一目でもみたいというヤジ馬もかなり多かったようである。ところでその湯川さんは肝心の原爆問題については完全にノーコメントでおし通した。記者団がしつこく質問すると『私は原子力については何も知らない。私のやっている学問はそんなものではない』と、いささか迷惑そうだった。湯川さんの原爆ノーコメントは何もいまにはじまったことではない。昭和二十四年に『中間子の理論』でノーベル賞をもらったとき、訪れた UP の記者にも『私は原爆については話したくない。私たちの話題はメソン（中間子）だ』と語っている。しかし一般の素人の考えでは米国のラビー教授（一九四四年度ノーベル物理学賞）が『湯川博士の新学説は第二次大戦に先立つ十年間の基本原子学説にもっとも重要な寄与をなすものだ』といっているとおり、ひろく原子力問題の最高の学者とみている。＜中略＞湯川さんがいくらノーコメントをつづけても国民は決して「湯川さんと原子力」を切り離して考えないだろう。（松田）」[36]　湯川秀樹博士が一般国民から原子力の権威であり、ビキ

---

指す。科学者らは反対したが、3 月 14 日に衆議院を通過。日本学術会議は平和利用限定で容認した。同じ 3 月 14 日には第五福竜丸が焼津港に帰港し、その 2 日後被爆が報道された。

ニ水爆の救世主であるとの期待を持たれていたことがわかる。

湯川には高知への出発前の京都でも、新聞社などから問いあわせがあったり、コメントを求められたりしたと思われる。原子力・ビキニ水爆について質問が出ること自体は、想定外のことではなかっただろう。京都でもそうであったように、湯川は一切答えなかった。だが、この湯川の対応は高知駅長室で質問した記者には意外だっただろう。京都と高知は違う。高知は太平洋に面し、遠洋漁業が盛んで、室戸や土佐清水や宿毛などの漁港からは多くの漁船員が水爆の海域に出向き、操業している。記者の知人、関係者、取材先にも当然漁業関係者がいるであろう。マスコミは県民を代表している。湯川は、高知が京都と違う地であり、漁業が大きな産業であるということに、どれだけ配慮できていたか。長旅の疲れもあり十分でなかったようだ。新聞記者（県民）の落胆ぶりが伝わる記事である。アメリカから帰って間もなく、そしてひたすら研究のみに専念していた働き盛りの湯川にとって、記者の言う「完全にノーコメント」[36]は無理もないことだっただろう。

この時の社会のビキニ水爆に関する関心の強さは、その年の8月1日に緊急出版された武谷三男(1911-2000)の本『死の灰』[37]の編集後記によく表されている。武谷は「編集後記」に、「ビキニの死の灰の事件は日本国民をあげての、否世界の大問題である。これほど全国民が頭をなやまし、その意見が一致したことは最近ないことである。本体が何かを知りたいという声に私はとりまかれていて、私への講演依頼は毎日行ってもしきれない位である」と記している。武谷のこの本は1954年8月1日に第1刷が出版され、8月15日には第2刷が出ていて大変な売れ行きであり、国民の知りたいという欲求の強さを示している。湯川の「私の研究外だ。この問題は答えられない。」[35]という記者団への回答は、明らかに「国民に背を向けた」と受け止められてもやむを得ない社会状況である。

湯川の宿泊先は坂本龍馬（1836（天保6）- 1867（慶應3））が生まれた家の近くの上町の城西館である。この高級旅館は天皇家の定宿として知られ、現在も繁盛している。当時は情報の入手は新聞かラジオに限られていた。湯川はビキニ事件と第五福竜丸の記事に目を通したことであろう。3月21日朝日新聞朝刊には「水爆被災・第一号か」という記事が出ている。

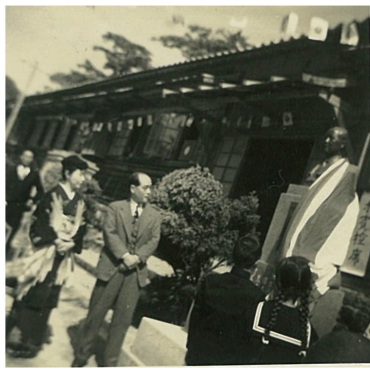

図3　高知県夜須小学校における湯川胸像除幕式に立ち会う湯川秀樹と澄子夫人。１９５４年３月２２日（浜田英子提供）[2]

翌日3月22日には湯川胸像除幕式のため夫妻で夜須町の夜須小学校に向かっている。この日3月22日の朝日新聞朝刊には「原・水爆実験の跡　米、既に四十六回」という記事が出ている。

湯川が核廃絶について発言をする方向にベクトルの向き変え、方針転換の決断をしたのは1954年3月22日夕方の一般市民向け講演の後から、夜の三高高知同窓会を経て、翌朝（3月23日）までとみられるので、以下に湯川の発言とその変化を詳細に追ってみたい。

## 5.2　3月22日午前　夜須小学校「湯川胸像除幕式」での発言

　高知新聞1954年3月6日[22]は、湯川の歓迎について沿道に小中学の生徒を集めて歓迎する計画があると次のように報じている。「"せめて顔でも······"湯川博士早くも引っ張りだこ　一九四九年ノーベル物理学賞に輝く湯川秀樹博士の講演会は各方面の前人気を呼び高知市中央公民館に聴講入場券の問合わせに来る人が一日いらい毎日五十名を数え、博士胸像の除幕式の行われる夜須小学校に至る沿道の小、中学校からは生徒を集めておくから顔だけ見せてもらいたいという要望も続々と届いている」

　筆者の父（大久保幹生、1915-1997）は当時高知市の中学校（昭和中学校：現在の城東中学校）の教師であった。高知市から夜須町の湯川銅像に向かう沿線の学校であり、これらの湯川歓迎計画や生徒の動員の話にはかかわっていたと思われるが、幼かった筆者には聞いた記憶は残っていない。父は南国市十市（阿戸）の石土神社近くにある兼業農家の生家から高知市まで通っていた。理科教師として、ノーベル物理学受賞者の湯川秀樹の高知訪問に大きな影響

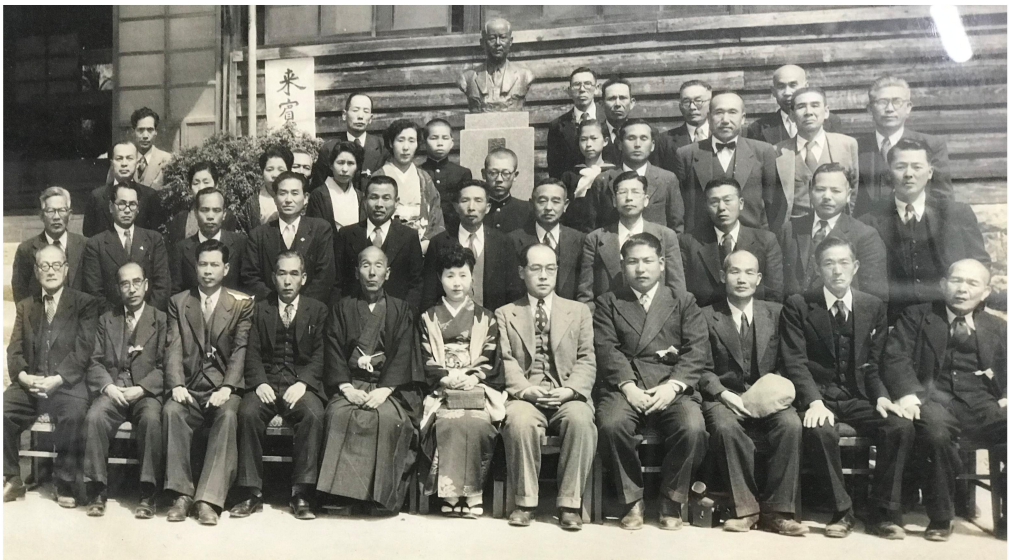

**図4**　高知県夜須小学校での湯川胸像除幕式後に学校正面玄関前で撮影された湯川秀樹夫妻と胸像設立関係者、学校教職員、来賓。前列中央：湯川秀樹、澄子夫人。右から2人目：末久愿（胸像設立賛同者）。湯川夫人の左：川村晴吉（夜須小学校PTA会長）。その左：近藤亘（夜須小学校校長）。2列目左端：野島憲（教育長）。左から5人目：横山楠壽（夜須町議）。右端：上田栄吉（学校医）。右から2人目：小谷武清（夜須町議）。3列目左から5人目：西内恵子（夜須小学校教諭）。6人目（湯川胸像の左隣）：高橋南海男（除幕した夜須小学校6年生）。右から5人目：春樹英子（除幕した夜須小学校6年生）。来賓者は胸にリボンを付けている。胸像台座に刻まれた「湯川秀樹先生像」の「湯」の字が見える。1954年3月22日、夜須小学校湯川胸像除幕式後、夜須小学校にて撮影（清藤禮次郎提供）[38]

を受けたようだ。学校での理科教育により一層専心する決意を強くしたようで、1、2年かけ先祖伝来の田畑・山林を整理し、湯川の高知訪問の3年後、生地から通勤にも便利で教育環境もよい高知城近く（小高坂）へ転居した。筆者も十市小学校（4年）から母（大久保政子、1921-2008）

の生家に近い高知市の小学校（小高坂小）へ転校（1956 年 10 月）となった。生涯の転機である。こうして筆者も、湯川高知訪問の影響を無意識ながらも受け、中学（城北中学）、高校（高知学芸高校）を経て、京都大学理学部物理学科に進むことになるから、筆者の運命に湯川の高知訪問は影響している。湯川が高知訪問をしていたことを筆者が知るのは、高知県立大学を定年退職後の 2018 年のことである。拙稿「湯川秀樹先生のはじめての胸像は何故高知に建てられたか」[1] を 2019 年に公表し、2022 年には 「湯川胸像除幕式への湯川秀樹先生の高知訪問: 生涯の転機」[2] を記すことになった。ふり返ってみると人生とは自分のあずかり知らぬところで、運命づけられている面もあるようだ。湯川と同じく筆者も 30 歳代の父も時代の流れの勢いのなかにあった。

　湯川夫妻は午前 10 時半、香美郡夜須小学校で行われた同博士の胸像除幕式に出席した。除幕式では湯川が挨拶を行った。夫人もその挨拶を聞いた。図 3 は湯川秀樹胸像が夜須小学校の 6 年生により除幕されるのを夫妻が見つめているものである。校舎には万国旗と思われる旗が飾られている。この写真は前稿 [2] で初めて掲載したものである。 教職員との集合写真を撮ったことは知られていたが、その存在は不明であった。その集合写真が発見された [38] ので図 4 に示す。湯川は除幕式のあとのスピーチで「きょうはこんなにりっぱな胸像をたてていただいて感慨無量です、私はあたり前の人間にすぎないが根気よく勉強をつづけただけのことです、この胸像がみなさんの役にたてば大変光栄です」[39] と述べた。（下線は筆者）「みなさんの役にたてば大変光栄です」という発言には、前日の高知駅での記者会見での「原子力」への質問にたいするつれない回答とは違って、人々の役に立ちたいという「湯川の思い」がにじみ出ている。

### 5.3　3 月 22 日午後　城山高校講演での発言

　22 日午後は夜須小学校から近い赤岡の城山高校で講演した [39]。この講演については文献 [38] で詳しく報告したのでそちらも参照されたい。講演の概要は次の通りである [39]。「城山高等学校における湯川博士の講演会は二十二日午後零時半から同校講堂で開き聴衆二千余名、場内外は立

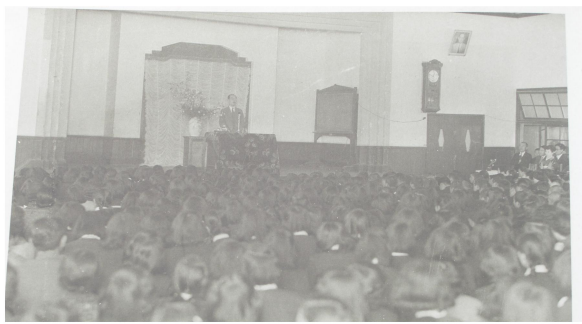

図 5　1954（昭和 29）年 3 月 22 日高知県赤岡の城山高校講堂で湯川秀樹（壇上）の講演を聞く高校生。壇上右に湯川澄子夫人と学校教員が見える（清藤禮次郎提供）[37]

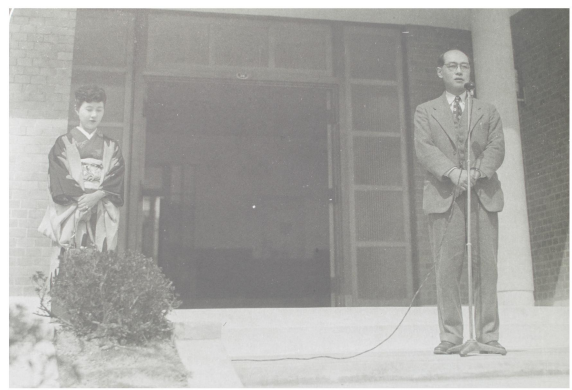

図 6　1954（昭和 29）年 3 月 22 日高知県城山高校講堂での講演に入れず湯川の講演を聞けなかった高校生のために玄関で改めて同じ講演をする湯川秀樹。壇上左に湯川夫人が見える（清藤禮次郎提供）[37]

すいの余地がない盛況ぶり、足立校長のあいさつについで博士は国吉 PTA 会長の紹介で登壇。私は幼い時から志を立ててこれを実行してきた。将来もこれで進む覚悟である、あながち偉人になるのがえらいのではない、皆様は志を立てて将来一つの研究に努力を傾け有能な人物になって<u>世界人類のため尽くしてほしい</u>と約四十分にわたって講演」[39]（下線は筆者）

「さらに玄関先で入場出来ない聴衆のため前とほぼ同様の講演を行い、終って澄子夫人とともに同校職員と記念撮影した」[39]。図 5 に講演の写真が載せられている。壇上に立っているのが湯川秀樹、湯川夫人は前方右に座って講演を聞いているのが見える。聴衆の高校生に女生徒が多いのは文献[38]で論じた。図 6 は講堂に入れなかった聴衆のために、外で立って同じ内容で 2 回目の講演をしている模様である。この時も湯川夫人は左側に立って聞いている。講演で「世界人類のため尽くしてほしい」と聴衆に向かって述べているが、自らの願いの吐露でもあろう。「私は幼い時から志を立ててこれを実行してきた。将来もこれで進む覚悟である」の言明は志を立てた通り将来も研究を続けていくと力強く宣言している。この思いはその日の夕方の高知市中央公民館での一般向け講演でも受け継がれ、より詳しく展開される。

### 5.4　3月22日午後　龍河洞の鍾乳洞を見物

湯川は城山高校での講演のあと、城山高校から近い念願の龍河洞見物に行った。龍河洞で見た蝙蝠のことが、翌日午前の学童向け講演で言及される[40]。湯川は講演原稿を準備するのが常であるが、ここでの体験を講演にすぐさま取り入れるのである。湯川夫人はかなりの距離の鍾乳洞内の歩行で疲れたとみえ、新聞は「きのう鍾乳洞（龍河洞のこと）へいったんで足が痛くて・・・」[41]と、翌日早速報道している。

### 5.5　3月22日夕方　高知市中央公民館一般市民向け講演での発言

夕方は、この高知訪問のメインイベント、一般市民向け講演である（図 7）。講演会の案内は 3 月 13 日にも高知新聞に再掲載されたが、3 月 22 日当日の高知新聞朝刊[35]に三たび、講演会の案内が 1 面に出た（図 8）。ここでも案内は「湯川博士の講演会」とあって、講演の演題はない。講演の演題がなくても講演会が成立する点からも、湯川の国民的な注目度の高さがうかがえる。前日の高知駅記者会見で講演内容について記者が質問するのももっともなことであり、湯川は「原子力についてはふれない、一般的な所見、科学者としての体験について述べる」と断言している。

夕方 6 時から始まった講演について、その聴衆の多さを新聞[42]

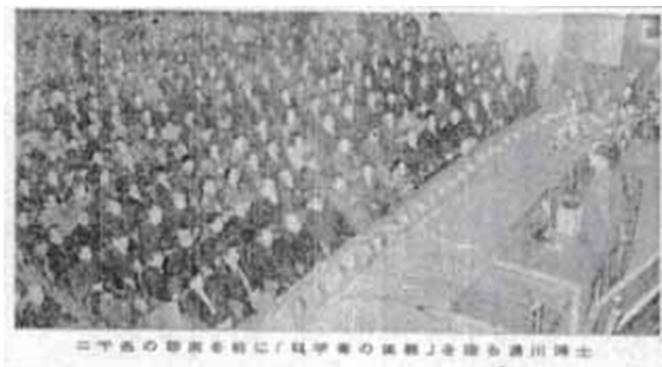

図7　１９５４年３月２２日高知市中央公民館における湯川博士講演会[42]

は「場外にあふれる聴衆」と見出しをつけ、「一九四九年ノーベル物理学賞を受けた京大教授湯川秀樹博士の講演会は県、市および県、市教育委員会、本社共催により二十二日午後六時から高知市中央公民館で開かれた。早くからつめかけた聴衆は会場前に列をつくり、約二千の人々が参集、福田本社社長の開会あいさつにひきつづき澄子夫人同伴で来場した博士の『科学者としての体験について』と題する講演が始められた。聴講券を手に入れることの出来なかった人たち約三百名も会場北側に設けられた場外聴取席に集まって夜気にもめげずマイクから流れる博士の声に熱心に耳を傾けるという姿も」[42]、と報じている。

講演の全文記録は残されていないが、講演要旨をみると湯川の苦悩がにじみ出ている。講演内容について新聞は「研究は知識欲のため 利己主義のそしりも甘受」[43]という見出しで報じている。元の文には速記のためか句読点が少なく、読みやすさのため適時句読点を挿入した。

「私が科学者になろうという考えをもった時期ははっきりしていない、私の父は地理学を勉強していたがそればかりでなく興味の範囲の非常に広い人であり、それらに何でもこる方で研究的にやる人だった。家にはあらゆる部門の書物や骨董が沢山あった。そういう環境で育ったので手当り次第あらゆる種類の書物を雑多に読んでいた。中学三年ごろアインシュタイン博士が来日せられ理論物理学というものが一般の社会の人の関心を呼んだが私自身にはわかる程の時期でもなく講演を聞きにいったこともなかった。しかしこういうことは自分では気がつかなかったが、それ以後の私の運命に非常な影響を与えていたのかも知れない。小学時代には算術、理科が割合好きであり、中学校、高校になるに従って化学、物理が得意であり興味もあった。しかし自分に一番適しているという判定はできていなかったが、大学には物理学科に首尾よく入ることができた。昔はアリストテレスのようにあらゆることに優れた人がいたが廿世紀の今日ではどの方面でも優れた人というものはない。何れかの部門に専心する人が偉い場合が多い。物理学にも実験物理学と理論物理学がある。私は理論物理学に優れているとは思っていないが実験物理学は得意でないということはわかった。またたわいもないことだが人との交渉がなくてもすむということも理論物理学に進んだ動機でもある。<u>一般には人</u>

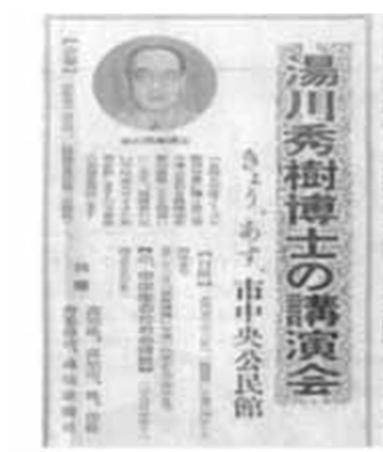

図8　1954年3月22日の湯川博士の講演の新聞案内[35]

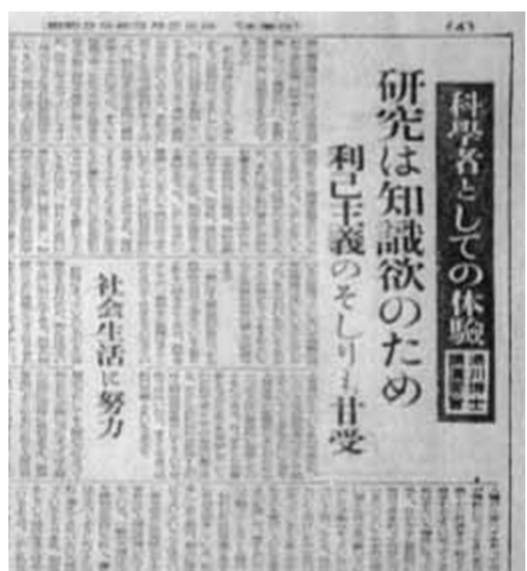

図9　湯川博士の3月22日の講演の概要を報じる新聞[43]

間生活に役立つから科学は大切だと考えていると思う。これも一理あることだが、科学者としての一番根本の動機はそうしたことではなく自分たちの生きている世界がどういうものであるか、いろいろな自然現象はどうして起るのかということを知りたいと思う人間本来の知識欲のあらわれから一つの方向に徹底的に勉強していくものである。はっきりした目標をもちどこまでも追求していくのが科学者としての本来の姿である。だからわれわれが研究するのは人間生活を豊かにするという目的ではないので、そうした点では利己主義ともいえるだろう。＜略＞研究によっていろいろのものを発見するわけだが、物には明るい面と暗い面が必ずある。科学の効能のために害毒が大きくなりそれをまぬかれることはむずかしい。原子力にしても人間の力によって作られたが使い方によっては人間の生活に非常に便利を与え、産業革命のようなものが起り人間生活が豊かになり幸福になることはわかっている。しかしこの原子力が人類を破滅にも導いていこうとしていることは一層われわれにはわかっている。こうしたことは世界の人に協力してもらわなくては危険をさけることはできない。科学者の知識欲によって発見されることが人類にとって望ましいことだけでなく、望ましくないこともあるので、そういう面で科学者は利己主義だとお叱りをうけるかも知れないが、そういうお叱りは甘受する。イヤなことは知らずにおこうと思っても他の国で研究されるから、いいことに使われるようになっても日本だけは手をつかねてみていなくてはいけない結果になる。また科学知識はだんだん進んでいってそれを逆向きにすることはできない。そしてそれは忘れていくということもできない。人類が大変ヘマをやって文明諸国が残骸になってもまた未文明国の誰かが自然界をよく知るという研究をするようになってくると思う。私はそういう運命になっていると思う。何かやると徹底的にしたい性分で骨を折って自分の力でしていきたいという本来的な傾向が幼い頃から今日まで続いている。私は大した才能はもっていないが自分でいいと思ったこと、自分でやりたいことだけしかやらない。自分のやっていることをとやかくいわれるのも嫌だし人のことを批判するのも嫌だ。要するに大へん利己主義でケシカラン人間だ。われわれは科学者であると同時に社会生活をしている。社会生活をしてゆく場合、他人に迷惑をかけてよいということはない。しかし稀な例外に芸術家らがあるが、それらの人達は世界で天才といわれている。一般の人達は天才という特別な生活に入れて、優れた人はなるべくそうしたワクの中に入れて考えてみたいという傾向を持っているようである。物理学者にもこういう傾向をあてはめて考えているようだ。しかし私は思うのである。物理学者のアインシュタインは天才に違いないが会えば普通以上の人という感じはしない。人間としての温かみのある健全な社会生活を送っている。私自身もやりたい仕事はやる。そして出来る限り人間としての立派な社会生活にも努力したいと反省し、そして努力しているつもりである。私が過去五年間くらいアメリカにいて日本へ帰った時、私から何か原子力の話をきこうとし、私がそれに触れると権威があるように思われているが、そうではないようである。私以上に原子力にくわしい人達は沢山いるのである。アインシュタイン博士は偉大な物理学者であるから原子力にくわしいか？そうでもない。興味を持っている方向も全然違うのである。昨年、京都で国際理論物理学会が開かれた際にも問題になったが、二十年前には研究の結果、数少い（ママ）二、三種類のもので自然界は成立っているといわれ、科学者を満足させる状態になっていたのが、戦後続々みつかり現在は数が

増えつつある時代で何とかしなくてはいけない。それを簡単にするためにいろいろ検討されたがこれという名案もでてこなかった。こんな状態であるからやめようにもやめられないのが学問である。<u>原子力</u>というものは原理的には現在わかっている。<u>利己的で無責任のようであるが原子力の問題は外の人にまかせておき私は本来的な傾向をおし通し自分のやりたいことに向かって進みたい</u>」（下線は筆者）

やや長く湯川の講演内容を紹介したが、これほどの赤裸々な苦悩の心境の吐露は、湯川はおそらく高知以外ではしなかったのではないだろうか。たとえば、地元京都の講演会でこのような赤裸々な心情の吐露を公衆の面前で行うとは、私の知る湯川先生の性格からは考えにくい。やはり、高知・土佐という土地が湯川をして語らしめたのではなかろうか。

講演内容について、新聞報道[43]は「湯川博士講演会　語る喜びと苦悩　感銘与えた"科学者の体験"」と題し、「人類を進歩に導くか破滅に落すかの岐路に立たされた科学者の宿命、苦悩あるいは研究の喜びを語る人間湯川博士の講演会は多くの意義を残して午後八時幕を閉じた。」と報じた。新聞記者が「人類を進歩に導くか破滅に落すかの岐路に立たされた科学者の宿命、苦悩」と書いているように、社会は湯川の講演を「原爆・水爆」の問題、具体的には「ビキニ水爆、第五福竜丸の被爆」と重ね合わせて受け取っている。

講演で湯川は、前日の高知駅での記者会見「<u>原子力についてはふれない</u>」との言明通り、「原子力」について詳細には触れてないが、「原子力」という言葉はこの要約だけでも6か所も出て来る、「科学者」は7か所、「科学」は3か所、「自然」は3か所、「利己」は3か所、「興味」は3か所、「物理」は13か所。当時の社会状況において、一般人が「原子力」で思い浮かべるのは、原子力の平和利用、原爆、水爆といった一般的なことではなく、「ビキニ水爆」である。湯川はビキニ水爆には直接触れなかった。間接的に「原子力にしても人間の力によって作られたが使い方によっては人間の生活に非常に便利を与え、産業革命のようなものが起り人間生活が豊かになり幸福になることはわかっている。しかしこの原子力が人類を破滅にも導いていこうとしていることは一層われわれにはわかっている。こうしたことは世界の人に協力してもらわなくては危険をさけることはできない」と述べ、ビキニ水爆・第五福竜丸事件に意図的に全く触れないよう留意していたように思われる。聴衆は直接的な言及を聞けるものと思い、湯川の一言一言に耳を澄ませて聞いている。

湯川の講演について、翌日の3月23日高知新聞コラム「小社会」[44]は、「原子力」に対する踏み込んだ発言が聞けなかったことについて、次のようにやや期待外れの感を淡々と記した。「国際政治の最重要課題としてだけではなく、第五福竜丸の"死の灰"事件以来、原子力問題は国民の身近な問題となってあらわれている折、博士にたいする国民の関心は深いが『私はみながいうほど偉人でもなければ立派な人間でもない』と語る謙虚な博士の科学者としての体験談こそ、県民の心耳には大きな響きを与えるものであろう」

また、3月23日の高知新聞夕刊コラム『話題』[36]で新聞記者の松田は湯川の講演について、「原子力」に触れず、国民・県民の期待に背を向けたことを、前日の高知駅での記者会見の内容も含めて、次のように厳しく批判した。先に引用したが、湯川講演に対する社会の反応の厳しさ

を示すものとして、改めて引用する。「湯川博士が来高した。ノーベル賞の湯川といえばこどもでも知っているせいか、高知駅は大した歓迎ぶりだった。ちょうどビキニの第五福竜丸事件で原子力への関心が一段と高まっている折だけに、原子物理学の権威、湯川博士の顔を一目でもみたいというヤジ馬もかなり多かったようである。ところでその湯川さんは肝心の原爆問題については完全にノーコメントでおし通した。記者団がしつこく質問すると『私は原子力については何も知らない。私のやっている学問はそんなものではない』と、いささか迷惑そうだった。湯川さんの原爆ノーコメントは何もいまにはじまったことではない。昭和24年に「中間子の理論」でノーベル賞をもらったとき、訪れたＵＰの記者にも『私は原爆については話したくない。私たちの話題はメソン（中間子）だ』と語っている。しかし一般の素人考えでは米国のラビー教授（1944年度ノーベル物理学賞）が『湯川博士の新学説は第二次世界大戦に先立つ十年間の基本原子学説にもっとも重要な寄与をなすものだ』といっているとおり、ひろく原子力問題の最高の学者とみている。湯川さんの沈黙は事をいやしくもせぬ学者的良心と解すべきであろうか。こんどの来高を機会に香美郡夜須小学校では博士の胸像除幕式を行うそうである。この学校でははじめ二宮尊徳の胸像を計画していたが、生徒やＰＴＡが相談のうえ湯川博士に決まったといわれる。昭和24年ごろには小学生の尊敬人物は湯川博士、エジソン、野口英世だった。＜略＞夜須小のこどもたちは湯川さんをどう考えているのだろう。恐ろしい原爆の恐怖を除いて原子力を平和へみちびいてくれる人、そんな漠然とした期待が学園の胸像となったのではないか。湯川さんがいくらノーコメントをつづけても国民は決して『湯川さんと原子力』を切り離して考えないだろう」。（3月23日夕刻は、湯川は高知を離れて京都に向かっており、この記事は読んでいないとみられる）

　だが、湯川の発言に微妙な変化がある点を見逃してはいけない。前日は「ビキニ被爆と政府の原子炉予算について」聞かれ、「私の研究外だ」[35]と断言し、一切の関与を拒否する姿勢を明確にしていた。この講演でも「ビキニ被爆」事件については一切言及していないので、首尾一貫しているように見える。だが、「原子力」という用語と「ビキニ事件」は新聞記者・聴衆にとって同義語に近い。湯川は一歩踏み込んだ発言をしたことになる。そのうえで、自分の研究は「利己主義」であるが本来研究とはそういうものだ、と強調している。「<u>私以上に原子力にくわしい人達は沢山いるのである</u>」との発言は、自身が原子力に一切関わりがないのではなく、より適任者が私以外にいるとの主張になっている。

　原子力に自分は関わりたくないという、かなり直截的な心境の吐露である。アインシュタインも原子力に詳しいわけではないとして、自身の主張の正当性を論理的に補強している。だが、後にそのアインシュタイン（A.Einstein、1879-1955）も、「ラッセル・アインシュタイン宣言」（1955年）で、核廃絶の運動の先頭に立つのである。湯川の講演の数か月後の8月に、「ビキニ事件」についての武谷三男の編集による啓蒙科学書『死の灰』（岩波新書）[37]が出版されている。これらの執筆者を見ると湯川よりもビキニ事件の放射能に詳しい武谷三男を筆頭とする専門家がおり、「<u>私以上に原子力にくわしい人達は沢山いるのである</u>」、という湯川の主張は根拠のあるものと言える。湯川が「<u>原子力の問題は外の人にまかせておき私は本来的な傾向をおし通し自分のやりたいことに向かって進みたい</u>」[43]というのは全くその通りである。世界の学問をリードする研究

者が専門でないことに引っ張り出され、研究ができなくなるのは、日本の損失であり、世界の損失でもある。

だが、湯川の主張は社会に受け入れられる状況ではなかった。「時代の勢い」であろう。江戸時代末期の「ええじゃないか」[45]という民衆の願望のうねり・勢いが討幕・明治維新へと展開したのを彷彿とさせる。実際、広島・長崎の原爆投下後に起こらなかった核兵器廃絶・原水爆禁止運動は、このビキニ事件から始まるのである。歴史的な巨視的な「集団運動のうねり」は事後的には認識できても、その始まりをその「構成体の一員」である個人が認識するのは困難である。2次元の平坦世界に住む生き物に3次元の世界が認識できないようなものだろう。

湯川夫人もこの講演を聞いている。湯川が研究者と水爆をめぐる自らの位置の認識について話すのは初めてであるから、湯川夫人もそれを聴衆と共に初めて聞き、研究以外のことで夫が苦悩しているのに驚いたことであろう。前年の夏に日本に帰るまで、湯川はアメリカで研究にのみ専念していたのである。

## 5.6　3月22日夜　三高同窓会

講演のあと、高知の三高同窓会「三高会」の宴会が開かれている。残念ながら、その性格上、記録は残されていないが、湯川の「原子力に向かう姿勢」に大きな変化を与えたと思われる。この宴会が開催されたことは、湯川夫人が翌日の桂浜散策で「ほんとに高知はいいところですわね　人情も景色も！　－それに酒もイイレネ、<u>みなさんお強いのにはおどろいた</u>、酒も少しなら頭脳の刺激になっていいもんですヨ」と発言しているのが、高知新聞に残されてる[46]。「みなさん」というのが三高同窓会の出席者であることは、高知新聞1954年3月6日の記事[22]に「在高の三高会では平田病院長、宮崎県農林部長、大岡高知赤十字病院長、浜田市建設部長、五十嵐高知大教授、二宮高知労働基準局長ほか二十名の同窓生が集まり二十三日午後六時から五台山荘で同窓のつどいを催すなど歓迎の方もひっぱりだこ」とあることからわかる。（注：記事では同窓会は23日とあるが、これは予定で、実際は22日に開かれた。23日夜は、湯川は京都への車中で高知に居ない）。 また、同窓生は20数名だとわかる。土佐の宴会は「おきゃく（お客）」と言われる。「少々飲む」とは「升々飲む」と言われるように、土佐には酒豪が多い。湯川夫人の「みなさんお強いのにはおどろいた」というのは、初めて土佐人と酒を飲んだ率直な印象で、さぞかし驚いたことだろう。筆者もその「土佐人」の一人であり、その通りだと思う。

湯川の高知訪問の仲介者として名前が挙がっているのは、新聞報道に出てくる高知労働基準局長の二宮と高知赤十字病院長の大岡の2名である。後に重要となるため、この二人について少し調べてみた。「二宮」とは二宮竜二は（にのみやりゅうじ）のことで、『高知年鑑　昭和29年版』[47]によると次のようにかなり詳しく記載されている。「高知労働基準局長、明治37年8月生まれ、大正14年三高文科甲類卒[48]、京大法科卒、出身　愛媛県　趣味　旅、野球　家族は伊十郎（父79）静子（妻44）弘子（長女20）」。令和の現在では見られないほど詳細な個人情報の記載である。三高同窓会の卒業生『会員名簿』[49]で調べてみると、p.335から始まる大正14年卒業生263人、うち文科甲類66人、のなかp.346に二宮竜二（京大法学部）と掲載されており[50]、三高から

京大法学部に進学したことが分かる。三高では湯川より1学年上で、湯川とともに三高生活を送っていることになる。国会図書館のデジタルコレクション『京都帝国大学一覧 昭和5年』[51] には、184コマ左のページより法学部の大正15年からの入学生の氏名が掲載されていて、二宮は昭和2年(1927年)に入学とある。湯川は大正12(1923)年4月に17歳で三高理科甲類入学、大正15(1926)年4月に19歳で京都帝国大学理学部物理学科入学、昭和4(1929)年3月22歳で卒業[54] とある。二宮は京都帝国大学で湯川より1学年下で、湯川と二宮の在学期間が重なっていることが分かる。

『高知年鑑 昭和30年版』[52] によると、二宮は京都、大阪、兵庫の労働基準局長を経て高知労働基準局長に着任し、その後滋賀労働基準局長に転任している。『高知労働基準局の40年』[53] によると、高知での在任期間は1953(昭和28)年1月16日―1954(昭和29)年3月16日となっており、在任は1年2か月と短い。湯川の高知訪問実現に労をとった1954年2月には高知にいたが、3月16日には離任し、湯川が高知を訪問した3月21日には滋賀県の労働局長に転任している。したがって、3月22日の三高同窓会の宴会に出席していない可能性もある。『高知労働基準局の40年』[53] には、「春秋叙勲」の中に勲五等双光旭日章として「二宮竜二元高知局長 昭和50年春」とあり、p.133には村田正廣の回顧録『思い出』にも出てきて、高知労働局長「三代目二宮竜二氏」として紹介されている。

次に、大岡について資料調べで判明したことを記す(図10に写真[55])。高知赤十字病院長大岡は『高知年鑑 昭和30年版』[56] によると大岡義秋で、「赤十字病院長兼外科医長、明治39年10月15日生まれ、京都府出身、京大医学部医学科卒 趣味 スポーツ 経歴：京大医学部副手、大阪北野病院外科長、市立長浜病院外科長」とあり、『高知年鑑昭和29年版』[57] によると「高知赤十字病院長、医学博士 家族 シゲ(母68)、静江(妻38)、秋朝(長男18) 宏記(二男8)」と、これも現在では記載されない事項まで詳しく記されている。3月23日に桂浜で湯川夫妻に同行し案内するのは、大岡義秋とその妻である大岡静江であることがわかる。三高同窓会の記録や写真は現在まで見つかっていないが、参加者の子孫に残され今後発見される可能性もあるため、ここに記載した。

三高同窓会卒業生の『会員名簿』[58] によると、大岡は大正15年理科乙類卒業生30名の名簿の3番目(p.363)に記載されている。国会図書館のデジタルコレクション『京都帝国大学一覧 昭和5年』[51] の194コマ右のページには、医学科の大正11年からの入学生の氏名が出身県とともに掲載されており、大岡は昭和2年(1927年)に入学したとある。これにより、大岡と二宮がともに昭和2年に京都帝国大学に入学したことが確認された。

湯川(小川秀樹)は明治40(1907)年1月23日生ま

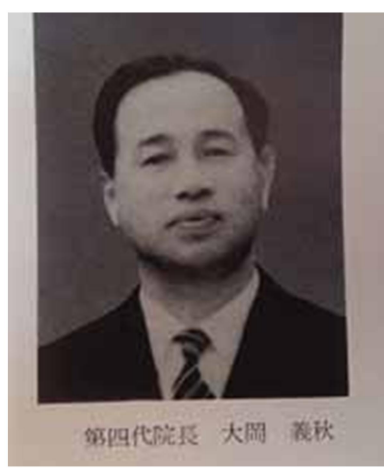

図10 湯川秀樹の京都一中、三高の親友の
　　高知赤十字病院第4代院長大岡義秋[55]

れであるため、大岡と湯川は中学（京都一中、京都府立京都第一中学校）で同学年である。湯川は1919（大正8）年4月に12歳で1年3組に入学、1923年（大正12年）3月に卒業し、同1923年4月（大正12年）に三高に入学している。三高同窓会卒業生の『会員名簿』(p. 358) [49] には、大正15年理科乙類卒業生104人の中に、湯川が「小川秀樹（京大理学部）」として記載されている。湯川（小川秀樹）は大岡と三高で理科類の学生として共に学び、同じ大正15年に卒業していることがわかる。湯川は京都帝大理学部へ進学し、大岡は京都帝大医学部に入学し、卒業した。1950年3月31日付けで、博士論文『経皮免疫に於ける表皮の存在意義に関する実験的研究』により、京都大学より医学博士の学位を授与されている [51]。

　昭和3年10月発行の三高同窓会の卒業生『会員名簿』[49] にはそれまでのすべての卒業生と教官名簿も載っている。湯川と同じクラスの卒業生に木村毅一がいる (p. 357)。湯川と大岡が教えを受けた物理学の教授、森総之助 (1876-1953) の名 [49] も p. 31 に載っている。三高における唯一の物理学の教官で、三高への就任年次は明治34年と記されている。森総之助は湯川が高知訪問に当たって「恩師故森総之助氏＝元三高校長に物理の手ほどきを受けたので高知は特になつかしい、妻も高知行きを希望している」と語り、平沢興 (1900-1989) 京都大学総長が湯川のノーベル賞受賞について「三高に物理学の先生で、森総之助という素晴らしい独創的な物理学の先生がおられたんです。この三氏（筆者注：湯川秀樹、朝永振一郎、江崎玲於奈 (1925-)）は、森先生の教え子です。つまり、ノーベル賞のもとは森総之助なんです」[59] と語っている。

　大岡義秋は京都で生まれ育ち、京都一中、第三高等学校、京都帝国大学に進み、医学と物理という違いはあるものの、同じ理科系の分野に進み、育ちや経歴が湯川とよく似ている。これらの経歴をみると、新聞が「中学時代の親友大岡高知赤十字病院長夫妻の案内で春光明るい砂浜を散策した」[46] と、京都人のふたりを「親友」と報じるのも頷ける。

　一方、大岡は生粋の京都人でありながら、高知赤十字病院に13年と長く務め、転勤後の他の勤務地では、自分はよく土佐人と間違えられた、と書いている。また、高知を「第二の故郷」とも書いており、「土佐人」としての側面、気質を有していることがうかがえる。高知での勤務では、楽しいことしか思い起こせない、と記し、まさに「土佐人と京都人の二重性」を持つ京都人である。このことを裏づける文を『創立50周年記念誌 高知赤十字病院』[60] に残しているので引用しておこう。「静岡にいるときもそうだったが、今の和歌山へ来てからも、私はよく高知出身のように思われたり言われたりしている。それだけよく私が高知時代のことを口にするからかも知れない。<u>事実、宴会などではよく"よさこい節"を故郷の歌のような気持ちで歌うことも多い</u>、私が高知赤十字病院に在任したのはまる13年とすこし、応召期間を引けば、私の勤務歴としてはもっとも永い高知である。その間、いろいろなことがあったが、今になって憶いだされるのは、愉快だったこと、楽しかったことばかりである。＜略＞ いずれにしても、高知の話は懐かしい。<u>第二の故郷などとよく言われるが、私にとって高知はそんなものかもしれない。</u>＜略＞その高知で、今でも季節になると憶いだすのは、高知城の裏山の公孫樹の道である。すれちがう人も殆どなく、散りしいた公孫樹の落葉を踏んで歩くと、明るい黄葉の中にも、何か物思わせるものがあって、私はひとりでこの道を歩くのが好きだった。それも20年ちかい昔のこと、あのあたりも開

発されただろうから、私の"哲学の道"も今はどうなっているのだろうか」

　湯川と長い交友があり、高知を第二の故郷と思う大岡義秋が中心となり、皆で土佐の皿鉢料理と酒を楽しみ、「三高同窓会」はさぞかし盛り上がったことだろう。土佐に生まれ育ち土佐で勤務した筆者は、職場やその他の関係で多くの宴会を経験したので、土佐の宴会がどう進むか想像するのは難しくない。1950年代当時、宴会は座敷で行われた。高知で客をもてなす宴会といえば、はりまや橋近くの明治から現在まで続く、名庭園の伝統と風格の料亭「得月楼」が筆頭に挙げられる。高知城真下にある講演会場の中央公民館からも近く便利で、おそらくここで行われた可能性が高い。得月楼は筆者も職場の宴会や客人を迎えての宴会で何度も利用してきたが、歴史を感じさせる料亭である。あるいは、高知城や中央公民館からは至近の鏡川河畔に位置する、土佐藩主山内容堂の屋敷跡に建つ由緒ある美しい庭園を残す昭和24(1949)年創業の三翠園だったかもしれない。三翠園は、西郷隆盛(1828-1877)が訪れ、幕末の1867年2月、山内容堂(1827-1872)と会見した所としても知られ、園内には「山内容堂・西郷南洲　会見の地」の立て札がある。

　いずれの場所であれ、湯川夫妻を歓迎するには十分なところで、主賓の湯川の隣には大岡義秋が座り、参加者は青年時代に返ったように、宴会は当時の学生コンパさながらに大いに盛り上がったことであろう。大岡の紹介で湯川夫妻の歓迎の宴会が始まり、湯川も挨拶の言葉を述べたことだろう。参加者は20人くらいなので、一人ひとりの一言もあったかもしれない。土佐では主賓に盃を持って挨拶に行くのがふつうである。この席は30年ほどはタイムスリップした三高時代であり、湯川秀樹は小川秀樹でもある。「小川君、ひさしぶりだなー。よく土佐へきた。まあ、いっぱい飲め」。こう言って盃を小川君に差し出す。小川秀樹が注がれた盃を飲み干す。土佐では、もらった盃を勢いよく飲んで、相手に酒を注ぎ返すのが礼儀である。相手は盃が返ってくるのを待っているのである。初見の挨拶である。返ってきた盃を飲み干すと、その盃を湯川夫人に渡し「奥様、ようこそ土佐へいらっしゃいました。」などと言いながら酒を注ぐ。湯川夫人も「ありがとうございます」などと言いながら飲み干すのである。「いいお味でした。ありがとうございます」などと言いながら盃を相手に返す。相手は盃が返るのを待っているのである。湯川夫人から返礼の盃に酒が注がれて飲み干すと、一通りの挨拶が終わる。「小川」ときに「湯川」かもしれないが、20人ほどの同窓生が湯川夫妻に同様に盃を捧げる。「升々」飲むというが、結構な量である。湯川夫人は「たまげた」ことであろう。翌日「みなさんお強いのにはおどろいた」[46]と発言していることには、真実味がこもっている。土佐の宴席では常道であり、あとは席を入り乱れて旧交を温めたことだろう。小川秀樹は何度も盃を貰ったことであろう。土佐人は酒席で天下国家を論じるのが好きである。湯川が講演で触れた、原子力や第五福竜丸の水爆被爆の話も出たことであろう。互いが酒の勢いで三高時代の若者に帰って大声で論ずれば、湯川も場の雰囲気に心地よさを感じたことだろう。土佐では宴会で箸拳という遊びをする。1950年、昭和の時代の宴会ではつきものであった。二人が手に箸を複数持ち、手を後ろに回し見えないようにして、二人の持っている箸の合計本数を当て合う遊びである。外れると罰であると同時に褒美とも言うべきか、二人の間におかれた盃の酒を飲むことになる。これを皆でやるのでますます盃を重ね、酒量は増える一方である。小川秀樹も土佐人気質の大岡やほかの同窓生たちとこの箸拳をやったの

だろうか、あるいは単に眺め興じただけであろうか。湯川は若い時から酒が強い。（筆者は湯川が若いころ宴席の女将であった祇園の女将から湯川の話の及んだ時『あの人は強いですね』と直接聞いたことがある）。こうして盃を延々と交わすことで、初対面同士でも100年来の友人のごとくになるのである。ましてや青春をともにした同窓生である。大岡が書いているように、土佐の宴会（この地では「お客」という）では、興がすすむと各人が銘々あるいは全体となって歌を歌う。大岡も「定番」の「よさこい節」を歌ったことであろう。この歌は、皆が手拍子を打ちながら歌い飲むのがふつうである。土佐の宴会と歌詞がよく合っている。京都人の大岡が土佐を第二の故郷と思うほど、気に入ったのもよくわかる。少し長いが雰囲気を解せるよう、歌詞を引こう。「♪① 土佐の高知の はりまや橋で 坊さんかんざし買うを見た ハアヨサコイヨサコイ／ ♪② 御畳瀬見せましょ 浦戸を開けて 月の名所は桂浜 ハアヨサコイヨサコイ／ ♪③ 土佐は良い国 南をうけて 薩摩おろしがそよそよと ハアヨサコイヨサコイ／ ♪④ 西に竜串 東に室戸 中の名所が 桂浜 ハアヨサコイヨサコイ／ ♪⑤ 思うて叶わにゃ 願かけなされ はやる安田の 神の峰 ハアヨサコイヨサコイ／ ♪⑥ 言うたちいかんちゃ おらんくの池にゃ 潮吹く魚が泳ぎより ハアヨサコイヨサコイ」。宴会中は一度ならず何回も歌われることもある。小川秀樹も湯川夫人も青年のころにかえって手拍子をうって口ずさんだことだろう。酔いが回ると、隣に座った大岡義秋と小川秀樹はたがいに名を呼び捨てながら、旧交を温めたことだろう。筆者も高知での在勤中、「土佐吉田会」という京都大学の同窓会に可能な限り出席した。最近は京都から現役の京大教授が来て最新の研究のお話を聞いてから懇親会・宴会に移り、旧交を温めながら親睦を深める。会の終わりは、三高寮歌「紅萌ゆる」や「琵琶湖周航歌」を肩を組みながら、歌い締めるが恒例である。湯川の歓迎同窓会でもおそらく宴会の締めには三高寮歌「紅萌ゆる」が歌われたのではないだろうか。「♪紅萌ゆる岡の花 早緑匂う岸の色 都の花にうそぶけば 月こそかかれ吉田山・・・／♪見よ洛陽の花霞 桜のもとのをのこらが いま逍遥に月白く 静かに照れり吉田山」。三高卒業生・京都大学卒業生にいまも吟じられ、青春時代の郷愁をさそう名歌である。46歳の湯川秀樹は10代の青春の小川秀樹にかえり、43歳の湯川夫人も若返って楽しいひと晩であったことだろう。湯川はこの一夜でかなり癒されたのであろう---。

　孤高な湯川が心を開いてはなしができる旧友・仲間である。20名と人数的にもみなとゆっくり話せる規模である。懐かしい人も多かったであろう。回顧談とともに天下国家、最近の話題も放談されたことだろう。湯川は童心・青春に帰り[61]、夜須小学校胸像除幕式で述べたように「人の役に立ちたい」との思いをさらに強くしたかもしれない。酒の好きな湯川は心置きなく土佐の銘酒と皿鉢料理を楽しむとともに、土佐の同窓の友人たちの龍馬を思わせるような大きな気持ちに背中をおされたであろう。苦悩の底にあった湯川秀樹、小川秀樹、に化学変化がはじまる。湯川夫人も同じ化学変化があったに違いない。翌朝の湯川の発言には、これから見る様に明らかな変化の兆しが現れる。

　この湯川秀樹歓迎の同窓会を書いていて、似た会合として、筆者は恩師の京都大学名誉教授の小林稔先生(1908-2001)が講演で高知を訪れたおり、歓迎の宴を小林研出身者や京都大学関係者と持ったことが思い起こされる。1970年代の昭和の時代であり、湯川の歓迎の宴を持った1950年

代の土佐のよき伝統は残っていたように思う。職場の同僚や友人とも毎月旬のものを味わうという宴を持っていた。今から思うと、月給の少なくない分がこの宴に費やされたようにも思うが、土佐の伝統でもある。小林稔先生の歓迎の宴も、湯川の件で述べたような雰囲気であった。筆者がこれを鮮明に覚えているのには理由がある。高知城下中心、鏡川畔の江戸時代の歴代藩主を祀る山内神社近くの会場の座敷での宴であった。小林稔は言うまでもなく湯川の直弟子であり、中間子論第4論文の共著者であり中間子理論建設に貢献した。湯川の三高・京大での同級生である朝永振一郎(1906-1979)とも親しかった[62]。小林稔を囲んで宴を楽しんでいたところ、女将が、電話が来ています、と呼び出しに来て、小林先生が中座した。当時は現在のようにスマホや携帯電話がなく、回線固定電話での呼び出しであった。朝永の著書をたくさん出版している出版社「みすず」からである。小林先生が座に戻られ、語られるには、編集者松井巻之助(1913-1984)からで、朝永先生が亡くなられたというものであった。一同暗澹たる気持ちでご冥福をお祈りし、朝永先生の回顧談に話が及んだ。筆者も朝永先生の話は直接聞いたことがあり、また名著で量子力学を勉強したので、沈痛な気持ちであった。湯川や小林の時代の学者は2025年の今日の大衆化し法人化された大学の時代と異なり、大学も学者も時間的にも精神的にもゆとりのあるゆったりした学究生活を送れていたように思われる。私が過ごした1960年代後半頃までは京都でも学者だけでなく、大学生も「学生さん」といわれ、京都市民に大切にされていたのを思いだす。1954年の湯川歓迎の同窓会の宴もそのような、敗戦後の色合いを残し物質的には恵まれないながらも、温もりのある人のつながりの強いよき時代であったように思われる。

### 5.7　3月23日午前　京都一中・三高時代の親友、大岡義秋・静江夫妻と桂浜清遊

翌朝、湯川夫妻は宿泊先の城西館をたち、念願の月の名所である桂浜へ出かけた[46]。同行の新聞記者は「"いい景色だネ"　湯川博士夫妻けさ桂浜見物」の見出しで次のように報じている。「来高中の湯川秀樹夫妻は二十三日の午前中を土佐の観光名所桂浜の清遊に過ごした。中学時代の親友、大岡高知赤十字病院長夫妻の案内で春光明るい砂浜を散策した湯川さんは遠く晴れわたった太平洋の水平線を眺めていささか感慨深そう、京都育ちの澄子夫人は波打ぎわで名物五色の石を拾って大よろこび。－　ほんとに高知はいいところですわね人情も景色も！　－それに酒もイイしネ、みなさんお強いのにはおどろいた、酒も少しなら頭脳の刺激になっていいもんですヨ　夫妻仲よく顔を見合わせて、はじめて見る土佐の風物に上キゲン、複雑な原子の世界から解放された人間湯川さんの、のどかな一ときだった」[46]

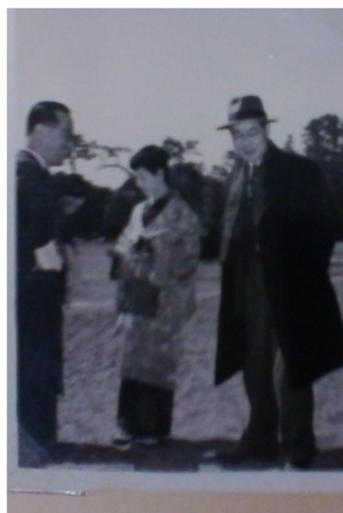

図11　1954年3月23日午前高知市の月の名所桂浜を清遊、大岡義秋（左）と湯川秀樹（右）、湯川澄子（中央）[63]

大岡夫妻が桂浜を案内しているが、同伴の夫人は先に出てきた大岡静江である。昨夜の久しぶりの同窓会で中学からの旧交を温めた二人である、大岡は桂浜でいろいろと湯川に説明したことだろう（図11、12［63］）。桂浜に入るとまず目にするのが高さ13．5メートルの本山白雲(1871（明治4）－1952（昭和27))作の坂本龍馬像である。高い台座の上にたつ龍馬像は見上げる必要があり、湯川も大岡と共に見上げたことであろう。昭和3年に有志の募金で建てられたものである。約250年続いた徳川幕府に終わりを告げ、万国公法によ

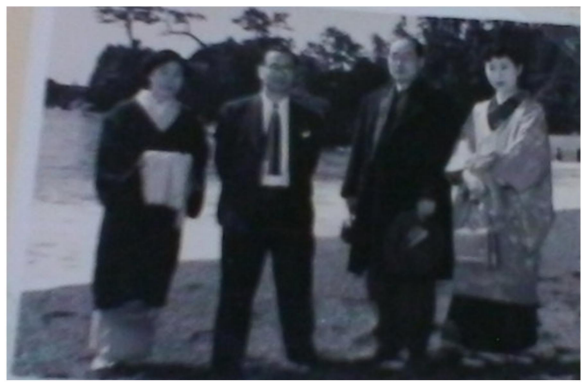

図12　1954年3月23日午前高知市の月の名所桂浜を清遊．左から大岡静江、大岡義秋、湯川秀樹、湯川澄子［63］

り世界と交わる開国日本の実現を夢見ながらも龍馬は一夜狂風の凶刃に京で倒れた。龍馬の世界に開かれた開国日本の実現を夢見る理想と思想は、土佐清水出身で1824年に漂流ののち米国の捕鯨船船長に救助され、米国で教育を受けた最初の日本人であるジョン万次郎（1827（文政10）－1898（明治31））から聞き取り、米国の文明を記した『漂巽紀畧』［64］を記した土佐の絵師・知識人である河田小龍（1824（文政7）－1898（明治31））［64］から受けたものである。ジョン万次郎は湯川に先立つこと100年前、1844年にMITの前身となる学校の一つであるバーレット・アカデミーに上級船員になるべく入学し、航海術、測量術、天文学、物理学、高等数学を学び、二等航海士の資格を取っている［65］［66］。ジョン万次郎が学んだ物理学の教科書は、オルムステッド著の教科書『AN INTRODUCTION TO NATURAL PHILOSOPHY』(物理学入門)（by Denison Olmsted, 1850)と思われる［66］。これはエール大学で使われていた当時の標準的教科書で、力学、静電気、レンズ、電磁場、音響など基礎物理学の内容を網羅している。大岡が桂浜で銅像を見ながら坂本龍馬について湯川に語ったことは確かだろうが、「土佐を第二の故郷」と思う大岡が龍馬に世界への目を開かせ討幕へ駆り立てたジョン万次郎の事はどれほど話しただろうかは知る由もない。いずれにしても、湯川は昨晩の三高同窓会での旧友との団欒に加えて、土佐の景勝地である桂浜を旧友の大岡と散策しながら気持ちを癒され、思いを新たにしたことだろう。新聞が報じる湯川夫人が浜辺で楽しんだ「五色の石拾い」［46］は、当時はよく行われ、「お土産」などとして売られてもいた。仁淀川などから海流で運ばれてくる小石などが自然の力で美しい五色の石になるようで、最近はこの桂浜名物の五色石も減ってきた。図11、図12の写真で、大岡夫妻・湯川夫妻の背景に写っている桂浜は、現在とは違って浜は波打ち際が近かった。筆者の記憶する1950年代の景色である。筆者は高知市出生であるが、幼少期は桂浜の対岸の種﨑に隣接する南国市十市（阿戸）の父の生家で育った。『土佐日記』の紀貫之（872-945）が都への帰路、正月に立ち寄った大湊とも比定され［69］、海の神の住吉神社とされ延喜式に出る格式高い石土神社があり、十市は和歌にも詠まれる歌枕で石土池はいまも残る。かつてその石土神社があった海上安全の四国巡礼三十三番札所禅師峰寺も近く、江戸期の文人、斎藤昌水［70］も記録に残している。桂浜は近い。大岡夫

妻・湯川夫妻の桂浜の写真を見て思いだすのは、十市小学校1年生の1954年のころ、遠足で桂浜に行き五色石拾いで遊んだことである。子供が夢中になる楽しい五色石拾いである。湯川夫人が五色の石拾いをし、子どものように楽しんだのは確かだろう。湯川夫妻は前夜の三高同窓会での土佐の人の気風に続き、桂浜では土佐の風光明媚に癒されたことだろう。湯川家が所蔵していた図11、図12の桂浜での写真は大岡義秋がのちに湯川に送ったものと思われる。大岡家の子孫や関係者宅にほかの写真を含め残されている可能性があり、今後これ以外の写真を含めて見つかることが期待される。

湯川が見つめた雄大な太平洋とアメリカを見つめる土佐の志士の銅像は、土佐の東端、室戸岬の先端には本山白雲作の中岡慎太郎（1838（天保9）－1867（慶応3））の像があり、ジョン万次郎が生まれた土佐の西端・土佐清水の足摺岬には、太平洋を見つめるコンパスと三角定規を手に持つ高さ3.6メートルの中浜万次郎の銅像が立ち、中央部の須崎・横波三里には土佐勤皇党を結成し土佐の志士を討幕へと導いた原寛山制作の武市半平太（1829（文政12）－1865（慶応元年））の大きな銅像が立っている。湯川は前日に夜須小学校で見た校庭に建つ「湯川秀樹胸像」を思い浮かべれば、その大きさと雄大さ、太平洋を望む龍馬像に感慨深いものがあったであろう。新聞記者は「湯川さんは遠く晴れわたった太平洋の水平線を眺めていささか感慨深そう」と書き、湯川の発する一言一言や表情を注視している。土佐人と思われるほどの大岡義秋は龍馬のことをきっと熱く語ったであろう。湯川が雄大な景色の桂浜での大岡との散策で癒され背中をおされたことは確かのようだ。直後の湯川の子供科学展での激励と学童向けの講演に、心境の変化がうかがえる。

### 5.8　3月23日午前　子供科学展入賞者の激励

高知訪問の最終日3月23日午前、湯川は桂浜から高知市中心部にもどり、子供科学展入賞者の激励を行った。これについて第3章で紹介したが、時系列として再録すると、新聞[18]は「湯川博士に激励頼む」との見出しで次のように報じている。「三月に来高する湯川秀樹博士に昨年十一月行った子供科学展に特選で入選した小、中学生の作品をみてもらい小さい科学者たちを激励してもらうことになった。同博士の中学、高校時代のクラスメート大岡高知赤十字病院長の話によると大正十一年冬、世界的学者アインシュタイン博士が来日したとき中学生であった湯川少年がその講演を聞き深い感銘を受け、この時から物理の世界に志をたてたということで大岡院長と片岡中央公民館長の肝いりで子供科学展特選組二十名を選んで来月の二十三日午前十時に行われる小、中学生に対する講演会終了後博士を囲んで懇談、博士のあとにつづく科学少年を本県からもだそうというわけ」

桂浜で一緒に散策した大岡義秋は湯川とふたりで子供科学展に入賞した子どもたちを励ました。その次の「学童むけ講話」の中で湯川はこのことについて「この講堂へ入る前に皆さんの研究された科学展を拝見致しましたが、これをみると昔の私達のころよりいまの方がずっと勉強の仕方や頭が進んでいるように思います」[40]と述べている。

理解を助けるためこの高知市の子供科学展についてすこし説明しておこう。子供科学展は、終

戦からまもなく1948年から始まる。（筆者も小学生（小高坂小学校）のとき「科学部」に属していて、この子供科学展に鉱物の研究で入賞したことがあり、子供科学展のことはよく記憶している）。この子供科学展はいまも高知科学未来館で「高知市小・中学生科学展覧会」として行われている。湯川の「激励」は1954年の3月の会のみであったが、現在では高知県で育った物理学者寺田寅彦が幼少・青年期を過ごした家、寺田寅彦記念館にある「寺田寅彦記念館友の会」が毎年「寺田寅彦賞」と「寺田寅彦記念館友の会会長賞」を授与するとともに協賛金を出して、子どもたちの「理科教育の推進と援助」を行っている。湯川の「激励」を受け継ぐ湯川と寺田を結ぶ不思議な因縁のある活動である。

　湯川と寺田は28歳ほども離れており、寺田は湯川が中間子論の論文を発表した1935年の大みそかに亡くなっている。湯川が寺田から直接の教えを受けた記録はないが、つながりはある。湯川は初めての日本人としてのノーベル賞受賞者であるが、それ以前にノーベル賞級の大きな仕事をした物理学者として挙げられるのは、1913年X線の結晶回折について先駆的な研究を行った寺田寅彦[71] [72]と1928年電子線の回折を示した菊池正士（1902-1974）[73]の二人である。寺田はイギリスの物理学者レイリー（J. W. S. Rayleigh、1842-1919）の音響学[74]の本でよく勉強し、1908年の学位論文の研究[75]「日本の竹製管楽器　尺八の音響学的研究」など、地震も含め波動現象に深い関心と専門的学識を持っていた。寺田寅彦はX線回折、原子の構造に関する研究・著作[76]に留まらず、原子核の構造に関する論文も書いている[77] [78]。寺田はラザフォード（E. Rutherford、1871-1937）が39才で、原子核の発見を発表した1911年に、英国マンチェスターで彼に面会している（4月28日金曜朝）[1]。以来、原子核に深い関心を持っていたとみられる。ラザフォードによる原子核発見の僅か8年後の1919年、日本人としては最初に原子核の構造を考察した世界的にも独創的・先駆的な原子核構造研究の論文を書いている[79]。潮汐力によるクラスター構造をもつ原子核構造の説明は同じ欧文誌に16年後に出る次世代・湯川の中間子核力による原子核の根本課題を解決するノーベル賞論文の登場を暗示するかのような先行的研究だ。寺田寅彦は日本における原子核構造の理論研究の先駆者でもある。寺田は日本におけるX線分光学の先駆者として、寅彦山脈ともいわれる多くの物理学の人材を残した[72]。菊池正士は東京帝大で寺田寅彦の教えを受けている[80]。菊池の研究はフランスの物理学者ド・ブロイ（Louis de Broglie、1892-1987）（1929年ノーベル物理学賞）による電子の波動性、物質波の考えを実験的に確認するものであるが、寺田によるX線の結晶回折の研究とは、回折による波動性の研究という点では共通性がある。湯川は大阪大学で中間子理論[81]を作り上げるとき、実質的に菊池研究室に属し、その自由な気風の研究討論やコロキウムなどで薫陶を受けていた[82]。湯川の中間子論[81]を導いた式は質量ゼロの光の波動方程式を質量のある粒子に拡張した波動場の方程式で、当時は適用範囲外と考えられていた原子核の世界に大胆に適用し解いたものである。

### 5.9　3月23日午前　学童むけ講話

　湯川は子供科学展入賞者の激励をしたあと、中央公民館で午前11時から1時間小中学生の学童

に対し講話をした。演題は新聞には出てないが『高知市中央公民館 26 年史』(昭和 52 年 5 月 31 日発行　高知市中央公民館)[83] ｐ．98 には「学問の仕方」とある。対象は「高知市内小中学生徒」とあり、参加者数は前日 22 日の「一般市民」対象の講演会と合わせて 2580 人と記録されている。高知市中央公民館の収容人数から 2 分すると、1200 人ほどの小中学生が参加したことになる。

　湯川が 1938 年、初めて徳島で中学生に講演したときは、ほぼ用意された原稿を読み上げるような完璧に準備された講演[84][85]であったが、今回の「学童むけ講話」の仕方は明らかに違う。話し言葉で思うところを話している。新聞は[40]「小学生のために　湯川博士のお話　早く特徴を伸ばそう　みなさん同様に私も生徒」の見出しで次のように書いている。「みなさんの知っている世界的な物理学者、ノーベル賞の湯川秀樹博士は二十三日市中央公民館でとくに本県の小、中学生のために次のようなお話をして下さいました。そして最後に『この話から一つでも参考になることがあれば幸いです』と結ばれました。集まった生徒たちは会場の都合から各校の代表者で多くのみなさんはお話を聞く機会に恵まれなかったようですから、ここにその要旨を述べてみましょう」。すこし長いが湯川の深層心理を理解するには重要なので、以下に講演内容[3]を見てみよう[40]。

　「私もみなさんと同じ生徒でありました。いまみなさんの受けている教育と四十年前に私の勉強したころとくらべると大変違っているように思います。この講堂へ入る前に皆さんの研究された科学展を拝見致しましたが、これをみると昔の私達のころよりいまの方がずっと勉強の仕方や頭が進んでいるように思います。これは理科の方面だけでなく全ての勉強もそうでしょう。私たちの小、中学生のころは大変楽であったように思われます。宿題がいまの皆さんのようになく、私は学校から帰るとカバンをほおり出して好きな本を読み、外で遊んでいた。幸か不幸か判らないし、どちらがよいとはいえませんが、これは

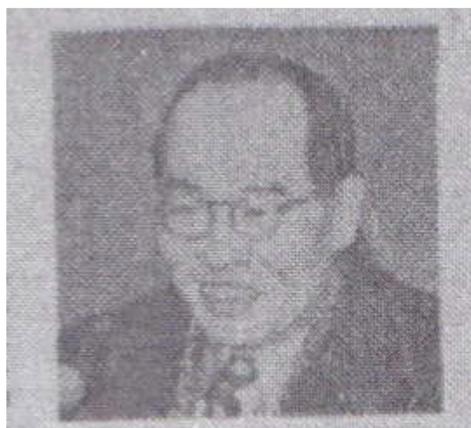

図１３　１９５４年３月23日高知市で小中学生に講話する湯川秀樹[88]

別として理科の方では確かに皆さんは進んでいるようです。私達小さいころは理科などが好きになる環境には恵まれていませんでした。算数は好きだったが立派な理科の実験設備もなく、どういう風に理科は勉強してよかったか判りませんでした。私達のころの子供達から理科方面のすぐれた人が出る環境と思われなかったのですが、ふりかえってみると私達の前後に沢山の科学者が集団的に出ているのは興味深いことです。昔は学問的にすぐれた人というとどの分野にもすぐれた人でしたが学問が進んでゆくに従い何でもできる人はなくなってしまいました。<u>物理でも私は</u>

---

[3] 筆者注：講演文には現在では人権擁護の見地から不適切と思われる表現がありますが、著者自身に差別的意図はなく、講演当時の時代背景また著者が故人であるという事情に鑑み原文通りとしました。

理論物理の方面で、この面では一応信用されても仮に生物の方で発言しても権威はないのです。このように学問はだんだん各専門にわかれていって、細かい研究をするようになりました。だから学者は一人だけとりだすといわばカタワです。これらのカタワの学者が集まって研究したものを持ちよって一つの体系が出来る、これが学問というものでしょう。私もそのカタワの一人であります。二十二日龍河洞に案内して頂きましたが、この洞の中には外の世界と違った動物が住んでいます。光が当たるとこの動物は絶滅するそうですが、人間もそれぞれの環境に応じて異なった人間が出来てくるでしょう。/みなさんの中には、これは得意だ、不得意だ、これをやりたいというものがあるでしょう。何よりも自分の特徴をみつけて伸してゆくことが大切なことです。ですからみなさんは特徴が自分で判らないときは父兄、先生たちと相談してみることが必要です。私も物理をやるまでいろいろ迷ってみました。迷ってみることがまた本当です。しかしこれだと決めてやり出すと変えないことが良いでしょう。三十年前私はやるべきことを決めて変えませんでした。学界にも洋服の流行があるように風潮というのがあります。この風潮に逆らうと思っても逆らえるものではありませんが、自分がやるべきことはやってきました。世間離れがしているともみえるでしょう。しかしこれは外からみていることでして、龍河洞のあの眼という概念のない動物にも、その世界があるものです。アメリカと日本では自からやり方、考え方が違います。しかしアメリカでも私は私なりのやり方を変えたことがありませんでした。日本人は外人と較べておとっているとよく問題にされる。しかし容易に決められないことです。なるほど学問では西洋は先進国でしたから沢山の学者が出ていますがこれでもって素質がおとっているとはいえません。アメリカで第二世、三世が大学に多く通っていますが、平均して成績がみな良い。成績が良いことがいろいろの先の問題からみて優秀だとは決められませんが、おとっている条件にはもちろんならないことでしょう。手先が器用なことは確かです。これもおとっている条件ではありません。日本人は物真似ばかりしているといわれます。これは実際にそうですが、明治以来外国の学問を取入れるに急であったためには仕方がなかったでしょう。真似をする半面にのみ込みが早いという長所があります。短い間に西洋に追いついたことはよかったですが、何でも真似をするのはかなしいことです。正しいことだと知ったらやるのです。しかし独りよがりは排すべきことです。日本人だけ、西洋人だけの科学というのはありません。科学の真理はどこでも一つ、西洋と日本で違っているというのは科学ではないでしょう。また日本人は熱し易くさめ易い性格がありますが、これは学問をしてゆくのに一番いけません。幼い時分からよく気をつけなければなりませんが、口でいうと何でもないが非常に難しいことです。何でも出来る人は何でもたいしたことはない人だと私は思います。偉い人とはどういう人をいうのか疑問ですが、いわゆる偉い人になるのが問題でないと思います。人にはそれぞれの特徴があります。全然ないという人も稀にはありますが、これは数少いことで、その特徴、自分にはどんな可能性があるのかを知って伸ばし世に役立つことが問題であります。多くの人は特徴があってそれを知っても伸ばさない、伸ばそうと思っても努力が足りない。こうして生甲斐のある生活を失っていってしまっているようであります。といって私は満足しているでしょうか。私はどうにかやっているが、いつも不満足です。それはまだまだ判らないことが沢山あるからです。この席で皆さんにお話しするよりも自分のや

るべきことをやらねばならぬと思っています。私はよい先生だと思っていません。それよりももっと判らないところを勉強したい。私は人に教える先生ではなく、みなさんと同様な生徒であるのです。この話から一つでも参考になることがあれば幸いです」（下線は筆者、読みやすくするため適時読点をいれた）

　湯川は自己を昨日の午後に龍河洞で見た「眼という概念のない動物」、蝙蝠にふれ、原子力に向かないことを暗示している。しかし、ここでは原子力に対するあからさまな拒絶、断固たる否定、ではないことに注目される。前日夜の一般向け講演とはトーンが子供向け講話とはいえ、やわらかである。「生物の方で発言しても権威はないのです」との発言も原子力には向かないことを暗示的に表現している。「断固たる拒否」から明らかに考え方の方向性（ベクトルの方向）が変わっている発言である。図13の講演している写真を見ても表情が穏やかに見える。湯川は和歌を好み詠んだ。和歌は暗喩の文学でもある。

## 5.10　3月23日午後　高知ロータリークラブ会員に講話

　湯川は正午から午後1時20分まで県商工会議所で高知ロータリークラブ会員に講話を行っているが[31]、講演記録は要旨を含めて現在までのところ見つかっていない。商工関係者のなかには漁業に関連する仕事の人もいると思われる。湯川が商工関係者にむけてどんな話をしたのか、ビキニ水爆についての質問が出なかったか、興味の持たれるところである。記録が見つかるのを待ちたい。

## 5.11　3月23日午後　専門家向け学術講演

　3月23日午後湯川は中央公民館で最後の講演である専門家向けの学術講演を大学教授、大学生、高校教諭を対象に行った。3月24日の新聞には次のように報じられている[41]。「学術講演会に臨んだ博士は『物質構成は従来永久不変のものといわれていたが、素粒子の発見によりこれまでのような研究の範囲では解決出来ず、非局所場ともいうべき無限の場を設けねばならぬ』と述べ、さらに原、水爆の将来について『素粒子の性質如何によってこれまでと異なったものへ成長することになろう。私は原子を平和の子としてあくまで育てる決心だ』と語った。」（下線は筆者）

　湯川は高知滞在中この講演ではじめて自分の専門の素粒子論研究について大学教授らに話した。「非局所場」について話しているので、当時湯川が取り組んでいた最先端の問題について話したことになる。「非局所場」については、湯川は先年の1953年9月に自身が会長となって戦後日本ではじめて開いた国際会議である「国際理論物理学会議」で「An Attempt at a Unified Theory of Elementary Particles（素粒子の統一理論の試み）[86]と題して講演している。同じタイトル「Attempts at a Unified Theory of Elementary Particles」で、その2か月前に、リンダウ（ドイツ）での第3回ノーベル賞受賞者会議でも講演している[87]。

　学術講演であるから、湯川は「研究外」[35]であると2日前に記者会見で公に断言した「原子力」にふれる義務はなかったと思われる。強制されるわけでもなかったのに、湯川が敢えて「原子力」にふれ、大きく踏み込んだ発言である「私は原子を平和の子としてあくまで育てる決心だ」

と「断言」したことは、湯川の気持ちに大きな変化が生じていることを示している。

「原子力」という言葉は使わずに「原子」という言葉を使っているが、当時は核兵器という言葉は今日のように使われていず、ビキニ水爆で放射能汚染されたマグロは「原子マグロ」と呼ばれている。「原子」のことばには、「原子力」、今日の「核」の意味合いが含まれて理解され使われている。同行の新聞記者は湯川の「原子力」に関する発言に注目して取材していると思われるので、湯川が「私は原子を平和の子としてあくまで育てる決心だ」と述べたのを聞き逃さず書き留めたのであろう。湯川は学術関係者の前とはいえ公衆の前で断言したのである。

一昨日の高知駅記者会見の「私の研究外だ」[35]、昨夕の一般向け講演での発言「<u>私以上に原子力にくわしい人達は沢山いる・・・原子力の問題は外の人にまかせておき私は本来的な傾向をおし通し自分のやりたいことに向かって進みたい</u>」[43]とは明らかにトーンが変化している。ベクトルの向きが負から正に変わったようである。暗示的な表現であるが、湯川が原子力にどう立ち向かっていくかの姿勢を表明したものだと受け止められよう。

3月23日にこの発言が飛び出すには、湯川は昨晩3月22日夜の三高同窓会後から3月23日の朝までに決意を固めたものと思われる。湯川がビキニ水爆事件後、おおぜいの人とまた親友・友人と互いに気を許し自由闊達にしゃべったのは高知訪問中でもこの三高同窓会が唯一である。3月23日には湯川の気持ちが大きく変化していることは明らかである。4日後の3月28日に毎日新聞への寄稿原稿を執筆するのもあり得ないことではなくなる。原稿の構想を練るのに4日ある。湯川は高知で決意したといえる。

筆者は湯川の決断の背景には湯川の背中を押す夫人の姿勢があったように思う。湯川夫人の自伝的著書『苦楽の園』[24]はよく知られていて、そこでは核兵器廃絶と世界連邦実現をめざす活動も触れられている。この著書には1954年の高知訪問のことは記されていない。湯川秀樹が大きな決断をする高知での湯川澄子夫人の発言は世に知られていない。湯川の「1954年3月決断」および科学者の妻として核廃絶・世界連邦実現に夫とともに後半生を捧げる「湯川スミ」の名前の誕生を理解するうえで、湯川「澄子」夫人の高知での発言には重要な示唆が含まれているように思われる。そこで、「湯川スミ」誕生まえの湯川澄子夫人の高知での発言を検証する。

### 5.12 3月23日午後 湯川澄子夫人の座談会での講演

これまで高知訪問中、湯川夫人は3月23日の午前までは秀樹と一緒に行動してきた。秀樹の講演もすべて聞いている。別行動をとるのは23日午後だけである。自身が講演するためである。その講演は山内神社の向かいにあるホテル三翠園で行われた。ここには土佐藩主の山内容堂の南御屋敷があり、

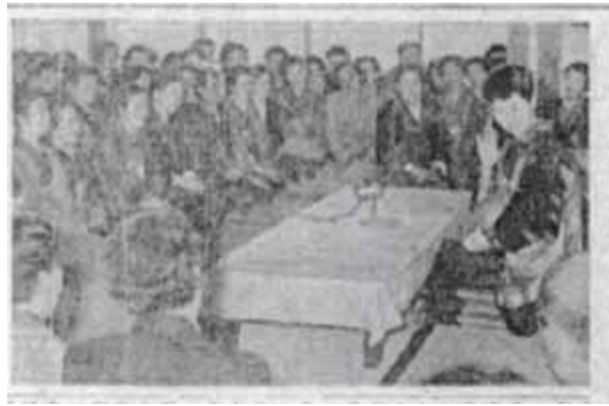

図14 1954年3月23日午後高知市三翠園での『湯川夫人を囲む座談会』で講演する湯川澄子夫人[41]

下屋長屋は美しい庭園とともに今も保存されている。講演の様子は、1954年3月24日の高知新聞に報じられている[41]。「湯川夫人を囲む座談会　扇さばきも鮮やかに　得意の"坂東流"もご披露」の見出しで報じた。「湯川博士夫人澄子さんは二十三日離高に先立ち午後一時半から三翠園で開かれた『湯川夫人を囲む座談会』に出席、会場からあふれるばかりにつめかけた一般家庭婦人約三百名との間にアメリカ生活あれこれを中心とした思い出話の花を咲かせたのち『きのう鍾乳洞（筆者　龍河洞のこと）へいったんで足が痛くて・・・』と笑いながら坂東流の見事な扇さばきで"京の四季"を軽くひとさし舞って並みいる奥さん連の目を見はらせた。"世界的科学者の妻"のお顔をひと目オガモウと押しかけたミーハー夫人族から『お花でお顔がみえません』などという声も飛び"スター湯川夫人"といったかっこう。夫人は終始ニコニコと意外なほど世間じみたザックバランで、しかも落着きのある話ぶりでアメリカ家庭の習慣、風俗などの長所や欠点を語ったのち『日本の婦人だって実に立派です。私も日本人であることを大変誇りに思っています。この美徳の上に<u>欧米の社交性を身につけたら世界一になる資格があると思います</u>』と結び『でもこれは男の方がいけないんだー』と付け加えて座を沸かせた。夫人は疲れた足を休ませるようにチョイチョイ横ひざになりながら『ご主人への心づくしの秘ケツは？』との質問に『頭を休ませるのが第一だと思ってこのごろはお酒を飲ませております』と答え、これまた満場を爆笑させた＝写真はなごやかな座談会場」（下線は筆者）

　以下にすこし長いが講演内容をもうすこし詳しく見てみよう[88]。湯川のその後の原子力への向かい方の変化についてのヒントが隠されている可能性がある。講演内容について、新聞の見出しは「日本夫人は社交性を　合理的な米国の都市生活　招かれても手土産なしで　大学に行く老婦人も」としている。

　「五年アメリカにいました。その間交際したのは学者の家庭が多かったのですが、渡米前に想像していたのとは大分違っていました。アメリカは機械文明が発達しているので都会に住めばアパートに電気洗濯機、乾燥機などの設備があるのでそれらを使用できるわけで実に時間を上手に使って合理的な生活をしています。何より遊んでいることを恥とするのでこどもが乳離れすると幼稚園に預け、自分は勤めにいくというのが普通の家庭の

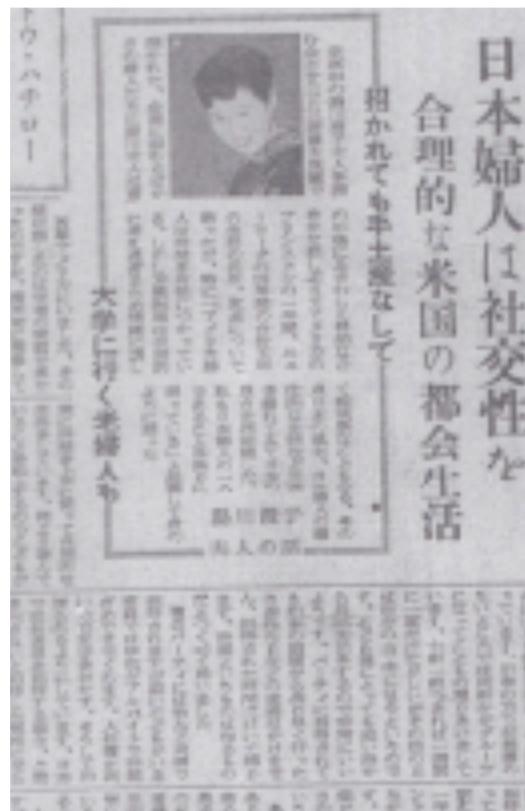

図15　1954年3月23日高知市三翠園での湯川澄子夫人の講演内容を報ずる新聞[88]

状態になっています。赤ん坊がいても授乳時間以外はカゴに入れてその時間は自分の時間として会に出たり教養を高めることに使っています。田舎の方で幼稚園のないところでは何軒かがグル

ープになってこどもの預りあいをしています。七軒一組であれば一週間に一度だけ忙しいがその他の日は自分の自由になるというのです。こども達にとっても幼い時から共同生活するので非常にいいようです。パーティに招待されても約束の時間から余り早く行ったり遅れたりなどの迷惑はかけません。招待された時間だけいて帰ります。時間というのは作るものだとつくづく思いました。

　夜のパーティには殆ど夫婦で招待されますが幼いこどもがいる家庭では学生がアルバイトで時間ぎめできてくれます。＜略＞日本ではお客様を招待する場合、ご飯を出さないと招待した気持ちにならず簡単にできないですが、午後にはお茶とお菓子だけのティ・パーティを、夜はお酒とつまみ物のカクテル・パーティをするので度々お客様を呼んでみんなと話をし知識を深めることができます。また招待されても無暗に手みやげを持っていかず、何かの機会に自分の家に招待します。機会がなければ一年中の分をクリスマスのプレゼントにして清算するので呼ばれたからと心配しなくていいようです。初対面の方がいる場合は主人側がその人に知っている限りの紹介をするので話題に困るということはありません。

　あちらの新聞は四十頁、五十頁あり日曜には二百頁にもなるので勤めを持つ主人達は読めないので奥さんが読んで食事の時の話題にのせるようになっています。ある老婦人がこども達が大きくなって手明きになったので、これから大学に勉強に行こうと思うといっているのを聞いて見習うべき点があると思いました。

　しかし学ぶべき点は沢山ありますが、それがかえって悪いことになることもあります。何ごとも余りに割切りすぎているため情緒が少ないこともその一つです。人間というものは無駄もあっていいものだと思いました。また都会では青いものや土を見ることができないので郊外に行くとこれはまた景色が大き過ぎて自動車で走れども走れども野原ばかりです。日本へ帰って汽車の中で日本はきれいだなと再認識しました。いたるところ山あり川ありで箱庭のように美しく感じました。また自分だけの家に住まうことも嬉しいことでした。日本の婦人は縁の下の力持ちだとよくいわれます。主人を毎日気持よく勤めに出し、こどもの教育に一生懸命になっていることは本当に立派なことです。最近若い人たちが自由をはき違えてこうした折角の美徳を失おうとしています。私も日本婦人の一人としてこうした奉仕は誇りに思っています。<u>しかし今までの婦人は余りに引込み過ぎていたために社交性に乏しいのです。これは男の人がいけないので、これからは女も一緒に出掛けることができるようにして欲しいものです。</u>日本婦人は社交術さえ身につけたら世界で一番の女性になることができると思います」（下線は筆者）

　昨日迄湯川夫人は主人とすべて行動を共にし、研究以外のところで社会の風圧に苦悩している学者・湯川の心内を知り驚いたことであろう。湯川はアメリカ滞在中の 5 年間、原爆やそれに関する話題は意識的に避け一切話をしなかったとのちに語っている [89]。湯川が「私は<u>原子を平和の子としてあくまで育てる決心</u>」を公衆に表明したころまでに、湯川夫人も「決心」の夫の背中を押していく「決意」をしたのではないだろうか。

　澄子夫人の講演を詳しく見たのは、その発言記録の存在が貴重であるだけでなく、随所に女性の社会進出にふれ、その後の「活動する湯川スミ」の誕生を暗示する発言が各所に見られるから

である。澄子夫人は聴衆、特に女性に、夫を支えながら社会へもっと出て活躍しよう、とアメリカ生活のいろいろ紹介しながら述べた。「夜のパーティには殆ど夫婦で招待され」「あちらの新聞は四十頁、五十頁あり日曜には二百頁にもなるので勤めを持つ主人達は読めないので奥さんが読んで食事の時の話題にのせ」「日本の婦人は縁の下の力持ちだとよくいわれます。主人を毎日気持よく勤めに出し、こどもの教育に一生懸命になっていることは本当に立派なことです」。「私も日本婦人の一人としてこうした奉仕は誇りに思っています。<u>しかし今までの婦人は余りに引込み過ぎていたために社交性に乏しいのです</u>」。これらの日本婦人への激励は自らへの激励でもあるかのようだ。澄子夫人の夫を支え新たな社会的活動を目指そうという決意が滲んでいるようにみえる。

### 5.13　3月23日午後　講演会終了後　念願の土佐犬を見る

　湯川は専門家むけの「学術講演」ですべての行事・講演を終えた。専門家にむけて専門的な講演で自らの研究である非局所場理論について話すことができたのは満足感を与えたことであろう。聴衆のなかには教え子の京都大学出身の者もいたと思われる。当時、高知大学文理学部助教授の物理学者の上田壽(1918-2009)[90][91]は京都大学理学部地球物理学科出身である。また、4歳下の高知県立高知女子大学講師で物理学担当の安藤敏幸[92]は京都大学理学部宇宙物理選科を卒業している[93]。会場でかつての京都大学での教え子にも出会ったであろう。湯川はすべての仕事をやりあげ満足感を覚えたのではないだろうか。

　会場の中央公民館を出ると湯川がぜひ見たいと思っていた念願の土佐犬が待ち構えていた。新聞[41]はこの模様を報じている。「湯川博士は高知市中央公民館午後一時半からの学術講演会に出席したのち、同三時会場前広場で市升形斎藤秀吉所有の土佐闘犬美濃号を見物した。これは『土佐に来たからにはぜひ闘犬とやらを一目！』という博士のたっての希望によるものといわれ初めて見る犬の化粧まわし姿に『ずいぶん大きいネ』とビックリしていた。」

　超ハードスケジュールのすべてを終え、あとは国鉄高知駅から鉄路で京都へ帰るのみである。予定では「三時五十六分高知駅発列車で離高する」となっている。その高知駅へ向かう。

### 5.14　3月23日午後　高知駅発国鉄列車で京都へ帰る
#### 5.14.1 高知駅で詩集を贈られる

　高知駅での見送りを報じた記事は見当たらなかったが、午前、湯川とともに桂浜を散策し、また子供科学展の入賞者を一緒に激励した大岡義秋も見送りの人々の中にあったであろうか。湯川夫妻は駅で、京都の湯川家の向かいに住み交際のあった知人の子どもで、高知市の小高坂小学校に通う六年生の坂野修一から自作の詩集を見送りの際に贈られた。その詩集を帰路の車中あるいは帰洛後読んだようだ。このことが4月5日の新聞[94]に「送った詩集に礼状　湯川博士夫人から一少年に」の見出しで報じられている。「先月来高した一九四九年度ノーベル賞受賞者、京大教授湯川秀樹博士夫妻が離高のとき高知駅まで見送った一人の少年があった。この少年は坂野修一君(11)＝小高坂小学六年、高知市山ノ端町＝といい、本紙にも紹介されたことのある少年詩人。坂野君にとって博士夫妻と親しくお話したあの日は、生涯の忘れられない瞬間だったが、さらに

喜びが重なった。このほど澄子夫人から便りが届いたのだ！坂野君の一家が終戦直後京都に住んでいたとき家の前に湯川博士の宅があった。坂野君はまだ幼なかったから知らないが、お祖母さんの兎美恵さんの話では澄子夫人と親しくつき合っていたそうで博士にも度々お会いしていたという。こんな関係でノーベル賞に輝く湯川博士は坂野君にとって偶像であるとともに身近な人だった。それだけに来高すると聞いてからは大喜びで小、中学生のための講演にはクジがはずれて行けなかったが高知市三翠園で行われた澄子夫人を囲む座談会にはおばあさんと一しょに出席、熱心にお話を聞いた。そしてお別れのときには高知駅まで見送って昨年暮はじめて二年生から四年生までの詩四十篇を集めて出版した詩集"あさ"を『汽車のなかででも読んで下さい』とお贈りした。ところがまさかと思った返事がこのほど舞込んだ。澄子夫人の筆で来高したときのことなどを書いたのち『本当に楽しく読ませていただきました』と感心して『こんな良い才能を』とまでいっている。すっかり感激した坂野君は『ボクも湯川さんのような人になるんだ』と大ハリキリ。この喜びを得意の詩に綴って夫妻にお贈りしたいとほおをかがやかしている。兎美恵さんの話　こんなに賞めていただいて・・・。私たちがお伺いしても博士まで出てこられて歓迎してくれます。修一も嬉しくて、嬉しくてたまらぬようです」

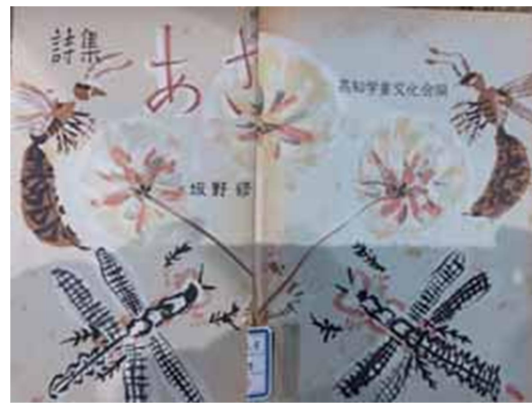

「本当に楽しく読ませていただきました」と書いているところからすると、夫人同様に湯川も「楽しく」感じたに違いない。そこでどんな詩が書かれていたのか探してみると、その坂野修一の詩集を見つけることができた（図16）[95]。小学校2年の時の詩14編、小学3年の時の詩14編、小学4年の時の詩16編の計44編である。このうちの11編は「小学生朝日」「こども文学」「子ども高新」「高知新聞文化欄」などに掲載されたものである。なかには高知市芸術祭で文芸大賞特選の詩「新しいクレパス」もある。前年の1953年6月30日には『詩集　あさ』の出版記念会が開かれ、高知新聞1953年5月27日夕刊が「30日の詩集　『あさ』の出版記念会」と報じ、6月1日夕刊が「『あさ』の出版記念会」と題し、記念会の様子を報じている。詩集の最後には、坂野修一の小高坂小学校4年で国語担任の安丸貞雄が「考える子供」と題する一文を寄せている。

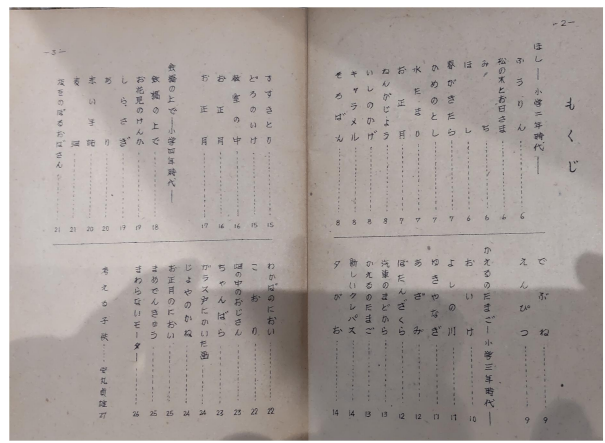

図１６　湯川夫妻が高知駅で見送り時坂野修一から贈られた『詩集　あさ』の美しい表紙（上）と目次（下）[95]

「坂野君が、このたび詩集を

出すことになったが、このようなすぐれた詩をものにすることができたのは、一体どういうわけであろうか。これは坂野君がいつも、ものごとを素直に見つめていく目をもち、考える子供であるからではなかろうか。ただ、みただけではだめだ、みたあとよく考える子どもでなければならない。こうした坂野君にこの詩が生まれるのもうなずけるわけである」。安丸貞雄先生は筆者が小高坂小学校で6年生の時にも6年生の担任としておられた。当時は1学級50人以上で今日の倍以上の人数であったが、戦後復興期で先生がたも非常に熱心であった。筆者の4年、5年生の時の担任の梅原咲子先生、安丸貞雄先生など2025年に創立100周年の伝統ある小高坂小学校には坂野さんを育むような教育環境があったようだ。湯川夫人が「こんな良い才能を」と書き、感心するのも道理である。高知市芸術祭特選の詩「あたらしいクレパス」を載せておこう。「新しいクレパスのふたをあけると/プーンとあぶらのにおいがして/あかるい色/しずんだ色が/いっぺんに目の中にとびこんできた　（小学生朝日）」。もう一篇「汽車の窓から」は「うすみどりの/やわらかそうな麦のほが/ちらっちらっとみえてかくれる/麦の手入れをしている/おひゃくしょうさんわ/ちらっちらっとみえてかくれる/とおい山の下で/すいしゃが/ことことんまわった」。湯川夫妻も車窓からこの情景をながめただろうか。この詩集を読んで、終戦直後、京都の自宅前に住んでいた幼い坂野くんの成長を思い起こしながら、また、今回の高知の旅を思いだしながら、心温まる思いで癒されたことだろう。歌人でもある湯川にとって詩集の題「あさ」はこれからの新しい「人生の転機」を考えると示唆的であったかも知れない---。

### 5.14.2 長い深夜の帰洛の旅路

　湯川夫妻にとって、高知駅から京都までの帰路は28ページの詩集を読むに十分なほどの時間がある。今日では、日本から地球の反対側英国ロンドンへ行くほどの長時間でハードな旅である。土讃線全体では100箇所を超すトンネルがある。当時、高松へ至るには黒煙を吐く蒸気機関車でそのトンネルを次々と抜け、四国山脈を越えなければならない。当日は春らしい天気で、夕方の春の美しい土佐の山野を車窓からながめながら山脈を越したであろう。当時の国鉄の時刻表を調べてみると土讃線高知駅発 15 時 56 分の 列車がある[96]。新聞に出ている湯川が乗車する列車で、普通2・3等列車である。高松駅着 21 時 45 分で、高松桟橋着は 21 時 49 分とある。四国山脈を超え、高松まで至るのにさえ6時間かかる。当時は21世紀の今と違って四国と本州をつなぐ橋はなかった。高松に着くと宇高航路に接続するが、時刻表[97]によると、高松桟橋 22 時 20 分発で宇野着 23 時 25 分の連絡船がある。湯川は深夜に高松桟橋から岡山県の宇野港桟橋まで渡った。この海の連絡船航路は霧が発生し、高知からの蒸気機関車による四国山脈越えと共に難所であった。（湯川夫妻が乗った翌年の1955年5月11には国鉄の連絡船同士が衝突し、紫雲丸に乗船していた修学旅行中の高知の南海中学校などの小中学生100人を含む168人が死亡し、第3代国鉄総裁長崎惣之助が2日後に責任を取って辞任している。この事故がのちの本州四国連絡橋の建設につながる）。湯川夫妻は無事に四国を離れ、本州に着いた。本州に上陸してからも京都駅まで6時間以上の長旅である。宇野から京都までは山陽線・東海道線の夜行列車であり[97]、宇野発は深夜0時08分で京都駅着は6時35分となっている。順調にいけば、湯川夫妻は3月24日

（水曜日）早朝6時35分に京都駅に着いたことになる。当時は夜汽車が普通だった時代とはいえ、相当に長い旅である。

　筆者は、先に触れた詩集の作者・坂野修一と同じ小高坂小学校を、彼より4年後に卒業しているが、湯川高知訪問の4年後（6年生のとき）の1958年秋の修学旅行で高知から高松の栗林公園に行ったことを記憶している。卒業アルバムで確かめると、高松は高知からの修学旅行先であるほどの遠隔地で、その時の蒸気機関車の旅から、湯川夫妻の四国越えの旅に思いを馳せることができる。

# 第6章　帰洛3月24日から「原子力と人類の転機」執筆の3月28日まで
## 6.1　湯川の思索と湯川の本来的信念

　列車の旅程を詳しく調べて書いたのには理由がある。湯川が京都に帰り思索する実質的な時間が何日あったかが重要であるからである。湯川の毎日新聞への寄稿原稿は3月30日朝刊に載る。湯川は高知滞在中は日記を書いていないが、帰洛後の28日からは日記を書いており、それが残されている[30]。

　「3月28日　日　雨　家に居て毎日新聞原稿「原子力と人類の転機」＜略＞

　　3月30日　火　晴　「毎日」朝刊に「原子力と人類の転機」第一面に出ている＜略＞」

帰洛した3月24日は高知でのハードな講演スケジュールと長旅の疲れもあり、休養日で旅行中の新聞に目を通すなどしたことであろう。表2に示されているように、帰宅の日からもビキニ水爆関連の新聞記事が連日出ている。湯川が京都に帰った3月24日から原稿執筆28日までは5日間である。湯川は3月28日には、毎日新聞に寄稿用のかなりの長さの原稿を仕上げている。休養日の24日と執筆日の3月28日を除くと原稿執筆までは25、26、27日の3日しかない。京都を留守にしていたのでそのほかの用事も多々積み重なっていただろう。「原子力と人類の転機」という、スケールの大きな題の原稿の構想には時間を要する。口頭でさえ「原子力」について触れるのを拒んでいた湯川が、文書で意見表明するとなると、これは相当な飛躍である。物理用語の「相転移」という言葉があてはまりそうだ。「清水の舞台から飛び降りる」ような飛躍である。京都に帰ってからも「私の研究外だ」[35]、「原子力の問題は外の人にまかせておき私は本来的な傾向をおし通し自分のやりたいことに向かって進みたい」[43]、と逡巡していては、原稿の構想どころではない。

　原子力について社会に文書で直接発言するには、湯川には相当の決断と相当の思索があったはずである。湯川は物事を深く、根本的かつ徹底的に考え、課題に取り組む性格である。中間子論の研究も、原子核を存在たらしめている根本的な原因を徹底的に追求した結果である。高知の「学術講演」で話した「非局所場」理論も、その後の「素領域理論」として「時空の問題として定式化」していく湯川にとって、単に「ビキニ水爆は遺憾だ」と新聞発表すればよいというような単純な問題ではない。「ビキニ水爆問題」を如何に自己のなかで位置づけ「理論的に定式化」するかが大きな課題であったはずだ。湯川は「ビキニ水爆問題」を、核廃絶と世界連邦の実現という後半生を捧げるようになる壮大な課題としてとらえる。高知で決意した湯川は、5日間熟考し、

結局、3月28日になって「原子力と人類の転機」として定式化し、毎日新聞社への寄稿となる。（この間のことは、後で触れるように、湯川は6月の三重県での中学校の校長先生向けの講演で吐露している）。

### 6.2 ビキニ水爆事件をめぐる社会情勢の変化

湯川が高知訪問を終え帰洛し毎日新聞寄稿原稿執筆までのビキニ水爆事件・原子力をめぐる社会情勢の変化を新聞で見てみよう。表2に3月24日から3月28日までの朝日新聞におけるビキニ水爆事件・原子力関係の記事の「見出し文」が載せられている。連日ビキニ水爆事件が報道されていることがわかる。

表2　湯川が帰洛し毎日新聞寄稿執筆までのビキニ水爆実験をめぐる新聞報道

| | 朝日新聞掲載日 | 新聞見出し |
|---|---|---|
| 1 | 1954年3月24日夕刊 | 損害補償は未定　コール原子力委員長談 |
| 2 | 1954年3月25日朝刊 | 科学者も驚かす　三月一日の水爆実験　ア大統領言明 |
| 3 | 1954年3月25日朝刊 | 原子灰で米、防空態勢再検討 |
| 4 | 1954年3月26日夕刊 | 水爆生産を促進　ストローズ原子力委員長証言 |
| 5 | 1954年3月27日朝刊 | ビキニ水爆と英の世論　日本の"恐怖"は正当 |
| 6 | 1954年3月27日朝刊 | 次の水爆実験に慎重　爆発力の警戒強化 |
| 7 | 1954年3月27日夕刊 | 水爆実験の映画を公開　四月七日に |
| 8 | 1954年3月27日夕刊 | 水爆問題に集中せん　労働党、質問を用意 |
| 9 | 1954年3月27日夕刊 | 放射能なし　米国へ到着のマグロ |
| 10 | 1954年3月28日夕刊 | 星の世界へも行ける　陸軍機関紙報道水爆の平和利用 |
| 11 | 1954年3月28日夕刊 | 来月二十二日か　第二次水爆実験 |

筆者は前稿[2]で、湯川胸像除幕式出席の湯川が、高知訪問を機に非核への社会的発言をする大きな決断をすることになったことを明らかにした。それでも稿のタイトルは『湯川胸像除幕式への湯川秀樹先生の高知訪問：生涯の転機』とし、「生涯の決断」とはしなかった。先に触れたように湯川の教えを受けた者として、その深い苦悩を公の場で論じたいとは思わなかった。前稿[2]の内容を朝日新聞は2022年4月に紹介し報じた[3][4]。ちょうどロシアによるウクライナ侵攻（2022年2月24日）がはじまったのと重なり、しかも核兵器使用の恫喝が国家と軍の最高責任者によって公然と語られていた。核戦争の危機が何度かあった20世紀の米ソ冷戦時代に逆戻りしたかのようであった。図17にその新聞報道（紙面版）[3]が載せられてい

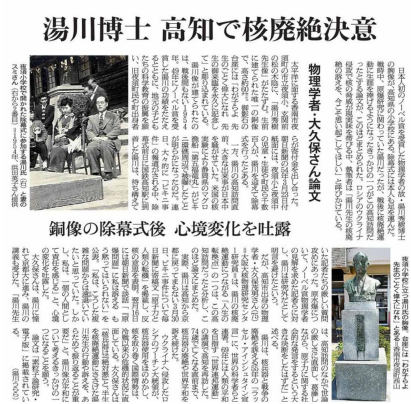

図17　論考[2]を「湯川博士　高知で核廃絶決意」と報ずる湯川うらら記者による2022年4月19日朝日新聞記事[3]

る。2022年4月27日の朝日新聞デジタル版[4]は「あの訪問先が湯川秀樹を決断させた　ビキニ事件直後の葛藤」と報じ、ロシアのウクライナ侵攻と核兵器使用の恐れにも言及した。筆者は湯川博士がなぜ高知で核廃絶を決意したかをより詳しく、社会のためにも、また湯川先生のためにも記す必要があると思うようになった。第1章で述べたように本稿を記す所以である。

　湯川がなぜ「原子力問題」で社会と関わることを頑なに拒否し続けてき、苦悩の末に1954年3月30日の毎日新聞での湯川声明にいたったか、は拙稿[2]で初めて明らかにされた。拙稿[2] 以前は、湯川の苦悩のことはまったく知られていなかった。湯川はそれまで学究生活に専念してきた。湯川が原子力とビキニ水爆は「私の研究外」[35]だと述べて、社会と関わることを断固拒否するのには当然ながら理由・論理がある。次の3点が挙げられよう。

　第1に、原子力は専門外、専門は中間子研究である。第2は、「自分でいいと思ったこと、自分でやりたいことだけしかやらない」[43]という、強い性格。第3は、本質論的理由として、「自分は原子力・社会・政治に関わりたくない」という信念・信条がある。

　第3に挙げた「信念」は非常に強く、湯川なりに合理的な根拠がある。第1に、戦前の原爆開発への関与についての自責の念と悔恨がある。関与は湯川の意図するところではなく、そのことは史料的にも立証されている[99]。湯川の海軍による原爆開発への関与はアメリカ側の証言でも消極的であったことが、政池明の研究[99]で明らかになったが、他方で原爆開発への関与については批判もある[100]。　第2は基礎研究といえども科学が政治と関わることのもつ危うさである。湯川夫人の自伝的著書[24]で記述されているように、湯川はアメリカ滞在中にアインシュタインから、米国大統領への原爆開発進言で政治的に関わり日本へ原爆が投下されてしまい罪もない日本人が多く殺されてしまった悔恨を直接聞いている。原爆開発を主導したオッペンハイマーが政治的嫌疑をかけられたことも知っている。しかし、第3の最大の理由は面白くてやめられない物理の研究を続けたいという学究的願望であろう。湯川には自分の研究人生のなかに「政治・社会」というパラメーターを入れたくないという信念は相当に強かった。アメリカでもオッペンハイマーらと広島・長崎の原爆について一切話したことがないと次のように言明している[89]。「私が米国にいたときには、全期間を通じて、原子力の応用についてだれかと話合ったということは、一つもない。<u>私は理論物理学者である。応用については、ちっとも興味がない</u>」（下線は筆者）長年強い思いをもって原子力・政治に関わることを避けてきた、その湯川がなぜ高知で「原子力・社会と関わる決意」をする気持ちになったのか。

## 第7章　湯川は葛藤を乗り越え社会の風圧をどう受け入れたのか
### 7.1　5段階の受容変化

　急激な変化が、社会でも個人でも、すぐには受け入れられがたいことはごく普通のことである。社会的・人間的な「慣性の法則」というべきものであろうか。湯川自身については、中間子論の発表でもそれがあてはまるようだ。1934年秋、11月頃東京大学物理教室における数物学会例会ではじめての中間子論の論文発表は「ほぼ教室一杯でしたが、内容についての討議も皆無で全く盛り上がりのない講演」だった、と小林稔が述懐し[101]、全くの無反応であった。無視されるの

ではなく、強力な拒絶反応も現れる。ニールス・ボーア(N. Bohr、1885-1962)は1937年の日本訪問時に「そんなに新粒子が好きか」と湯川粒子の予言に対して全く否定的であった[62]。パウリ(W. Pauli、1900-1958)の弟子のシュトッケルベルク(E. Stueckelberg、1905-1984)は湯川と同じ頃に似た研究をしていたが、パウリに否定され潰されたという[102]。「独創的な研究はすぐには受け入れられない」というのはよく知られているが、避けがたい科学者の宿命でもある。激変を受容するには緩和時間を要する。 物理学の慣性の法則はひとの心にも当てはまるかのようだ。

湯川は長く拒絶してきた「原子力」をどう受容したのか。急激な変化がどう受け入れられるか、その過程がもっともよく研究されている典型的な例は、「死の宣告」を受け入れる過程である。ひとはだれも子どもの事故による急死や親兄弟の急死などといった突然の急激な変化を受け入れるのは難しい。死という最も受け入れがたい事実の受容過程には、次の5段階があるという[103]。第1段階：否定。「そんなはずはない」という拒絶。湯川は京都でも報道陣に回答を拒否しただろうし、高知到着時の高知駅記者会見では「研究外だ」[35]と断固否定し、自分は関係ないという態度をとっている。第2段階：怒り。湯川の3月22日夜の中央公民館での一般向け講演は、「どうして私が？原子力にはほかに適した人がいる」のに自分に回ってくることへの「怒り」である。第3段階：取り引き。「もしも死の前に・・・ができるならば、避けられないという死の延期を願い、神との契約を求める」段階という。湯川が研究を続けたい、「私は本来的な傾向をおし通し自分のやりたいことに向かって進みたい」。「私以上に原子力にくわしい人達は沢山いるのである」[43]という発言は、研究は何としても続けたいと主張し、「第3段階」の「取り引き」的側面の発露かもしれない。第4段階：抑鬱。「しかたがない」という段階。湯川の学童講話での「外の他の世界を知らない鍾乳洞の蝙蝠」の話は、消極的受容の始まりではないか。否定することや不満を述べることはせず、次の段階に入っている。再度引用すると、「<u>物理でも私は理論物理の方面で、この面では一応信用されても仮に生物の方で発言しても権威はないのです。</u>＜略＞二十二日は龍河洞に案内して頂きましたが、この洞の中には<u>外の世界と違った動物が住んでいます。光が当たるとこの動物は絶滅するそうです</u>・・・」。湯川は光が当たると絶滅する動物など、いろいろたとえて自らを託し、諦念の心境の段階に入っている。研究者としての「死」：「<u>光が当たるとこの動物は絶滅するそうです</u>」は自己を動物に託した隠喩とはいえ、きわめて強い表現だ。子ども相手だから直接的な気持ちを吐露できたのかもしれない。第5段階：受容。「避けられない死を悟る。静かな沈黙、平和な休息、死を超えた希望」。湯川の3月23日午後の学術講演での「私は原子を平和の子としてあくまで育てる決心だ」の決意表明は、原子力と関わることを受け入れることを、覚悟したことの心の吐露と受け止められよう。こうして、湯川は高知で「原子力への関与と核廃絶の決意」を固めたといえる。受容過程はつねに5段階をとるとは限らず、ある段階が飛んだりする場合もあるという[103]。

**7.2 湯川が高知で決断できた4つの理由**

最大の謎はなぜ湯川が高知滞在中の短期間にこの5段階の受容過程をへて、決意を固めることが出来たかである。筆者の[2]の論考が発表されるまでは、毎日新聞声明にいたるまでの湯川の中

での深い苦悩や葛藤は知られていなかった。湯川の決意の背を押した要因として次の4点が考えられる。

　第1は三高同窓会への出席である。ここでは高校時代の旧友と再会し、青年時代にかえった率直な交流ができた。湯川が高知滞在中におおぜいの人と気をゆるして直接的に話をしたのは、この時だけである。特に京都1中からの親友である大岡義秋との再会・会話が大きかったと思われる。湯川秀樹ではなく小川秀樹として三高の友人と話すことが出来た。大岡は1月には京都におもむき湯川から高知訪問の了解取り付け、3月21日には高知駅では出迎えを行い、3月22日には三高同窓会で懇談し、3月23日には桂浜でともに夫婦で散策し、子供科学展入賞者の表彰といったように、滞在中の湯川とかなりの時間を共にした。旧友大岡義秋と話す湯川は、「湯川秀樹」というよりおそらく、「小川秀樹」ではなかったか。旧友大岡との再会・交流は、「小川秀樹」の背を押し、気持ちや心境に影響を与えたことだろう。

　第2は湯川夫人の澄子の支えであろう。澄子夫人は最終日の3月23日午後の自らを囲む座談会のとき以外は、夫とすべての行動を共にしている。湯川の3月22日夕方の2000人余の聴衆を前にした一般向け講演で、夫の深い苦悩を公開の場で初めて聞き、仰天したことだろう。今まで夫人は湯川の研究を支えてきたし、そのことを自負し誇りに思っていることは自伝[24]からよくわかる。湯川は養子であり、湯川家が財政的にも湯川の活動を支えてきたことは知られている。京都・知恩院にある「湯川家」の墓所にある湯川スミによる「湯川頌徳碑」[104]にある「湯川家は湯川城主の末裔。玄羊医学博士は初代湯川胃腸病院長。専門書を著述出版。秀樹を迎え物心ともに後援・・・」という記述には、夫人の湯川を生涯献身的に支えてきた自負がうかがえる。そこにはノーベル賞につながった物理学研究だけでなく、核兵器廃絶の運動も含まれている。

　夫人の「酒も少しなら頭脳の刺激になっていいもんですヨ」[46]などいった発言は、細部まで湯川の活動に気を配っていたことを示す。上述の5段階の受容過程は当事者だけでなく家族にも当てはまるという。おそらく夫人も、「夫の決意」を支える決意を同じく高知で固めたのではないか。湯川夫人はその後「湯川スミ」として、核廃絶運動で夫を支えていく。

　高知訪問以降と思われる湯川「澄子」から「湯川スミ」への名前の転換は、夫の原子力・核廃絶に関わる決意を共に支持・支援しようとする「夫人の決意」と関連しているのかもしれない。「湯川澄子」から「湯川スミ」にいつ頃いかなる理由で転換したのかは、筆者の知る限りではよく知られてない。これを明らかにすることは、夫人の「転機」がいつ頃であったかを知るうえで欠かせない。そこで、新聞などをたどりいつ頃変わったのか調べてみた。1954年の高知訪問時は「湯川澄子」と新聞に

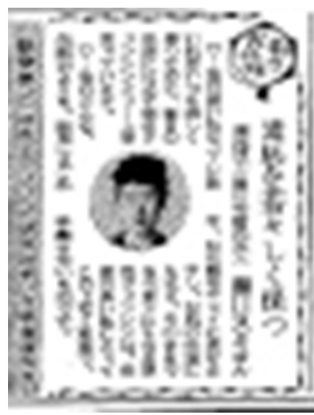

図18　読売新聞1958年4月27日[110]の化粧品広告『私のお化粧』に登場する「湯川スミ」

出てくるが、1949年1月1日読売新聞[105]には「澄子」ではなく、「すみ」が出ている。「世界

の檜舞台へ科学日本」と題する記事でアメリカ・プリンストン大学の湯川の近況を報じ、そのなかで「わが国原子物理学の権威京大理学部教授湯川秀樹（四二）同**すみ**（三九）夫人」、とはじめて「すみ」の呼称を使用している。アメリカではそう呼ばれていたので自然なことである。読売新聞は1952年9月9日[106]で、「湯川博士近く渡米」の見出しで、「帰国中の湯川秀樹博士は廿一日からコロンビア大学の講義が始まるため十一日午後一時六分京都発"はと"で上京、・・・十八、九日ごろ**スミ夫人**同伴渡米する」、と「スミ夫人」としている。ところで、同じ米国出発の読売新聞1952年9月18日の記事では、「二年ぶりで七月に帰国した湯川秀樹博士は来秋の国際物理学会議の打合わせや病気の長男春洋君（一九）の見舞などをかね約二ケ月を故国に送ったが、十七日午後八時卅五分ノースウェスト機で夫人**すみさん**（四二）とともに羽田を出発、三たびアメリカに向かった」として、「すみ」と平仮名表記である[107]。1953年湯川が米国から日本に帰ってから開かれた国際理論物理学会議について、同じ読売新聞（1953年9月17日）[108]は、「あすから本会議京都に移った国際理論物理会議」の見出しで「すみ子」と次のように報じている。「藤岡組織委員長以下日本側関係者の涙ぐましい努力・・・レセプション、観光には湯川博士夫人**すみ子さん**を委員長とする教授夫人団があたる」。高知訪問以前では、呼称は統一されてないが、「すみ、スミ、すみこ」といった表記が出てきており、いずれも夫と行動を共にする国際的な記事の場面である。

　ところが、高知訪問以降は呼称が明らかに変わっている。1956年になると、読売新聞は6月15日[109]「湯川夫妻欧州へ出発」の見出しで、「原子力委員、京大教授湯川秀樹博士はスミ夫人とともに十四日夜八時五十分羽田発KLM機でジュネーヴに向かった」。1958年4月27日の読売新聞[110]では、化粧品の広告に湯川夫人が載っている。広告であるにもかかわらず、新聞記事のような見出しで、「私のお化粧　素肌を若々しく保つ　理学博士湯川秀樹夫人　湯川スミさん」とあり、記事の左には広告会社の宣伝文句「新発売！『明色シリンシン』プラスチック新容器入り」とゴシックの大活字で網掛けされている。見出しにある「湯川スミ」も一段と大きい活字でしかもゴシックである。本文の書きだし部分は、「肌の弱い私は、アレ易い肌をいつも美しく保つために、毎日の洗顔には必ず明色クリンシンクリームを使っています。他のどんな洗顔料よりも、簡単によごれや、お化粧をキレイに落とせます・・・」とあり、記事の中央には「湯川スミ」の写真が載せられている（図17）。

| | 表3　湯川夫人の新聞にみる名前の変遷 | | | |
|---|---|---|---|---|
| 1 | 新聞記事出現日 | 記事内容 | 新聞名 | 湯川夫人　表記 |
| 2 | 1949年1月1日 | 渡米秀樹記事 | 読売新聞 | すみ |
| 3 | 1952年9月9日 | 渡米秀樹記事 | 読売新聞 | スミ |
| 4 | 1953年9月17日 | 学会秀樹記事 | 読売新聞 | すみ子 |
| 5 | 1953年9月18日 | 学会秀樹記事 | 読売新聞 | すみ |
| 6 | 1954年3月21日 | 高知訪問　同伴 | 高知新聞 | 澄子 |
| 7 | 1958年4月17日 | 広告、単独 | 読売新聞 | 湯川スミ |
| 8 | 1963年7月17日 | 世界連邦、単独 | 朝日新聞 | 湯川スミ |

いままでの報道記事で湯川夫人が出てきたのは、湯川秀樹と外国への渡航や学会等の記事中である。大衆向けの広告記事に「湯川スミ」という名で単独で出て来るのは、その名が宣伝力をもつほど国内で定着していることを示す。こうして高知訪問後、1958年4月よりかなり前に、読売新聞の1956年6月15日[109]「湯川スミ」報道のころまでには、湯川夫人は自ら「湯川スミ」に改名し、世間に受け入れられていることを示す。表3で名前の変遷を時間軸での整理をすると、高知訪問後4年間に公の場で自ら名のる名前が、本名「湯川澄子」から「湯川スミ」に変わり、湯川夫人の「活動の実体」も"相転移"していることがはっきりわかる。朝日新聞に「湯川スミ」という名がでてくるのは1963年7月17日朝刊が初めてであり、「世界連邦建設同盟（会長、<u>湯川スミ</u>氏）」[111]と出てくる。

　高知訪問の1954年3月23日から読売新聞報道の1956年6月15日の間に、湯川夫人は自ら名のる公式名を「湯川澄子」から「湯川スミ」へ変えた。湯川夫人の「生涯の転機」である。この「転換」は、湯川秀樹の「原子力と人類の転機」の執筆から核廃絶に向けての運動開始の時期と符合している。その間の「1955年」は7月に湯川が「ラッセル・アインシュタイン宣言」に署名し、8月には広島で第1回原水爆禁止世界大会が開かれ、核兵器廃絶運動が本格的に始まる年である。高知で「生涯の転機」を覚悟する湯川には、澄子夫人の背中を押す姿勢を、以心伝心で気づいていたではないだろうか。

　湯川の決断をあと押しした第3はおそらく土佐の風土と人の気風であろう。幕末・明治維新で土佐の志士は日本の政治体制の劇的転換に大きな役割を果たした。土佐には時代を切り開く進取の気風がある。湯川は土佐を講演と観光で各地をめぐりいろいろな人に巡り合い・酒食をともにして土佐の気風に触れた。親友の大岡義秋は「私はよく高知出身のように思われたり言われ」[55]「第二の故郷などとよく言われるが、私にとって高知はそんなものかもしれない」[55]という人物である。湯川は大岡から土佐の気風を感じ影響を受けたであろう。土佐人は、坂本龍馬、板垣退助（1837（天保8）- 1919（大正8））、中江兆民（1847（弘化4）-1901（明治34））、植木枝盛（1857（安政4）- 1892（明治25））などを挙げるまでもなく、天下国家を論じるのが好きであり、行動にも移す。「小川君、第五福竜丸はどうぜよ。政府はアメリカべったりで腰抜けぜよ。」同窓生の勝手な天下論国家論に湯川も興じただろうか。湯川が友人・旧友との宴会を通じて従来の頑な態度を変えるきっかけになったことは発言の変化から確かだろう。同窓会での宴会後の翌朝、桂浜ですがすがしい気持ちで見上げた巨大な坂本龍馬像も、湯川の背中を後押ししたことだろう。土佐は天下を変えてきたが、この風土と気風も湯川に影響を与えたであろう。湯川は高知で決意した。湯川が高知訪問せず京都に閉じこもっていたとしたならば、湯川の「心境の変化と決断」はもっと遅くなっていたかもしれない。その意味で、土佐が湯川を変えたともいえよう。土佐の同窓生、民衆が新生湯川を生み出したともいえるかもしれない。朝日新聞は記事で「湯川博士　高知で核廃絶決意」[3]「あの訪問先が湯川秀樹を決断させた　ビキニ事件直後の葛藤」[4]と報道したが（図17）、湯川が高知で「生涯の転機」の決断の覚悟をしたのは間違いない。

　だが、上記の3つの理由のほかに、最も重要な第4の理由は、大局的で大胆な決断ができる湯川の資質や人間性であろう。「生涯の転機」となるような判断・決断をほぼ1日足らずのうちに行

うことができる人はそれほどはいない。少なくとも凡人には出来ることではない。湯川にはそのような資質があった。湯川は「原子力」問題では、「核兵器は絶対悪」と規定し、廃絶にむけ世界連邦の構想を提唱・定式化して生涯の課題と設定する一方で、物理学・素粒子論の研究では非局所場理論を発展させ壮大な素領域理論へと展開していく。2つの展開は原子核の核力の原因として大胆な中間子論を構想し、理論的に定式化し展開したのと構図は基本的にそれほど変わらず、論理的に一貫性がある。湯川は「核廃絶」を生涯の課題とすることを決断した。湯川の優れているのは他人に促されるのではなく、自分で考え、自分で決断し、しかも高知訪問中の短い間に行ったことである。通常の人がたやすくまねできない偉人の業である。そこには湯川が幼少期・青年期に親しんだ漢籍、とくに荘子（BC369頃-BC286頃）の逍遥編にでてくる鵬を思わせるような大きな世界観があるように思える[98]。

## 第8章「原子力と人類の転機」執筆から核廃絶運動へ

　1954年3月30日に毎日新聞寄稿：「原子力と人類の転機」（図19）[112]で、湯川はこう述べた。「原子力の猛獣はもはや飼い主の手でも完全に制御できない凶暴性を発揮しはじめた」「原子力の脅威から人類が自己を守るという目的は、他のどの目的よりも上位におかれるべき」。格調高い核廃絶の必要性の定式化であり、湯川は「私は科学者であるがゆえに、原子力対人類という問題を、より真剣に考えるべき責任を感ずる。私は日本人であるがゆえに、この問題をより身近かに感ぜざるをえない。しかしそれは私が人類の一員としてこの問題を考えるということと、決して矛盾してはいないと信ずるものである」と自らが関わっていくことを宣言した。

　毎日新聞の寄稿文が公表されると、湯川はすぐさま、社会の激流に引き込まれていく。4月2日には「金　晴＜略＞午後　国会　自由党総務会で　原子力について話す」と日記に記している[30]。

　2週間ほどのちの4月15日には、湯川は湯川記念館で新聞記者の単独取材に応じ、ビキニ水爆にたいする率直な反対の気持ちを吐露する。朝日新聞はスクープ記事4月16日夕刊[89]でこれを報じ「『原爆問題』に私は訴える　＝もう黙ってはいられない＝」との見出しで湯川の談話を報じた（図20）。毎日新聞寄稿文とは比較にならない人間湯川の感性的な意思表示である。1か月弱前の3月21日夕刻の高知駅到着時の記者会見での発言「これ（筆者：ビキニ被爆と政府の原子炉予算）

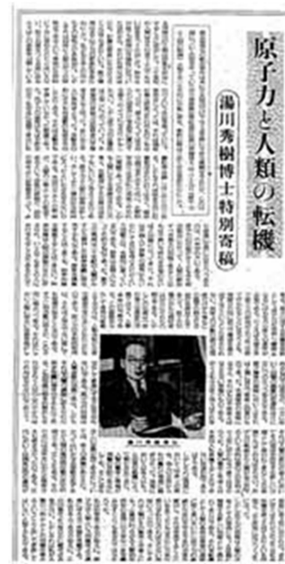

図19　1954年3月30日毎日新聞朝刊に発表された湯川秀樹博士の原子力にたいするはじめての社会的発言である「原子力と人類の転換」の記事[10]

についてはご承知のように一般に言明したことがなく全く関知しないところで私の研究外だ。この問題は答えられない」[35]（下線は筆者）とは大きな転換である。新聞記者、しかもアメリカの

通信社の東京支局長を京都の湯川記念館（京都大学基礎物理学研究所）に呼んで単独会見で語ったところに、世界にむけて自己の新たな決断を発信する湯川の固い決意がうかがえる。ニューヨーク・タイムズ[113]は4月15日京都発の記事として日本を代表する最も著名な科学者がアメリカのビキニ水爆実験に反対の態度を表明したと速報した。アメリカ占領軍による支配から日本が独立して3年もたっていないことに留意したい。アメリカ軍が京都占領中に京都大学理学部構内をアメリカ軍のジープが我が物顔に駆けるのを見て、湯川は次の和歌を詠んでいる：「わくらばに音立ててジープすぎゆきぬ銀杏並木をひとり歩めば」[114]。湯川は京都で占領軍に査問も受けている。同じ理学部の先輩教授であった原子核物理学の荒勝文策（1890－1973）は自ら製作したサイクロトロンをアメリカ軍に破壊され、研究資料もすべて没収された[99]。湯川はそんな辛い思いもおさえて、研究に打ち込んだ。湯川が日本に帰国したのは1953年7月16日（木曜）午後6時半であるから、つい10か月ほど前まで5年間もアメリカに招かれ滞在し研究生活を送ってきたアメリカに、公然と反対することは大変勇気のいることである。日本政府さえアメリカにまともに抗議すらできない時である。湯川のなみなみならぬ決意の強さがみえる。湯川はこうしたことをふくめ高知で決意し、3月28日までに「原子力と人類の転機」と定式化した。アメリカの通信社の東京支局長との単独会見でビキニ水爆被害者への率直な心情の吐露のあと、湯川は約1か月後の5月16日には第2回全国PTA大会（富山市）で「科学と人間－原子力問題と関連して－」と題して講演している[115]。

専門の物理学・素粒子論の研究では、5月26日自ら所長を務める京都大学基礎物理学研究所での基礎理論討論会第2回例会で「非局所場の理論について」の研究発表している[116]。6月23日には、同じく京都大学基礎物理学研究所での基礎理論討論会第3回例会で、「四次元量子化」と題して講演している[117]。アメリカから帰国後も非局所場理論の研究に精力的に取り組んでいる。

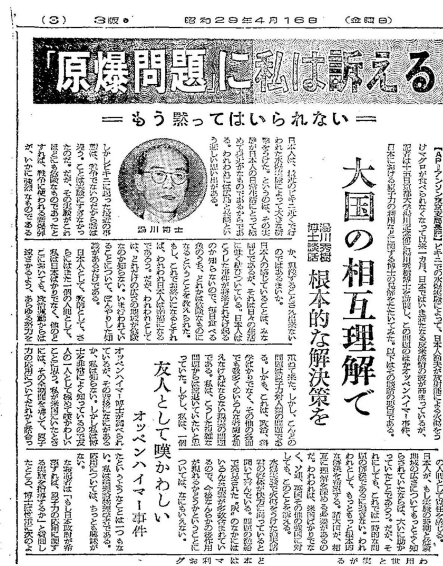

図20　朝日新聞（1954年4月16日）掲載の湯川秀樹の談話：「原爆問題」に私は訴える　＝もう黙ってはいられない[89]。

ビキニ水爆実験後の講演として、9月には「私の人生観の変遷」[118]と題し、三重県での中学校長の集まりで講演し、ビキニ水爆・第五福竜丸被爆からの「人生観の変遷」を述べている。湯川とほぼ同世代の学校の先生を対象の話ということもあってか、3月23日高知で学童に語った語り口と似ているところがあり、素朴な心境の変化・人生観の変化を吐露している。「現在の段階は、ここで一度深く反省をしなければならないような時期にきている。我々の一人々々が色々反省しなければならないような時期にきている。そういうふうに思うのであります。私自身も自分の経験に基づいて反省してきたのであります。」[119]と述べ、講演の結びで「反省」という言葉を繰り返し、「心境の変化・人生観の変化」を吐露して

いる。高知での葛藤についてはまったく触れていない。長い間の苦悩を自らのうちに閉じ込めているようだ。湯川は高知訪問のことをその事実を含め一切記録に残さず講演でも日記でも一切触れなかった。勝海舟（1823－1899）は「孤峰碧旻に秀ず」[120]と漢詩「偶感」に残したが、孤高の湯川の一面を見る思いがする。このころ日記につぎのように書いている[30]。「9月23日　木　晴　暑　＜略＞ビキニ死の灰の被害者久保山愛吉氏死去の報あり　新聞記者ら夜おそくまで押しかける」とある。

　高知で決意を固めた湯川は3月28日の「原子力と人類の転機」の原稿の執筆まで構想を練ったようだ。中間子論や素領域理論の構築[121][122]を見ても、湯川は、原子力やビキニ問題に断片的にまた個別的・現象論的対応や談話で済ませる性格ではない。原子力やビキニ問題を自己の中で理論的にどう位置づけるか、ビキニ問題の本質は何か、原子力の問題の本質は何か、自分はその本質論的課題にどう関わることができるのか、また関わるべきか、高知から帰洛の3月24日から執筆の28日まで4日程しかないが、湯川は中間子論に劣らない生涯をかける沈潜をしたであろう。広島・長崎の原爆投下、原子力の問題、ビキニ水爆問題、第五福竜丸被爆問題を現象論的・個別的対応でなく、本質論的課題としてどう位置づけるかである。湯川は結局「人類の転機」として位置づけた。「寄稿文」の最後で湯川が記したように、勿論その一員たる自己の転機でもある。その意味で3月28日の寄稿原稿の起草は湯川のその後の「核兵器絶対悪」、完全廃絶、世界連邦構想へと発展進化させられる「原点」である。自己の方向を定めた湯川は世界に向かって決然とビキニ水爆反対と漁民への同情を外国通信社の記者に表明し、実践へと踏み出していく。湯川のこのような深遠な考えが、駅頭記者会見で表明され理解されることは望むべくもなく、「研究外だ」[35]と拒否するほかないという、対応と心情は理解されるべき側面がある。孤高な人間湯川の「深淵なる苦悩」を理解できた人はいただろうか。おそらく湯川の性格をよく知る旧友の大岡義秋と澄子夫人は語らずとも湯川を理解したのではないか。

　湯川の決断は画期的なものであった。1955年7月核兵器廃絶を訴えるラッセル・アインシュタイン宣言が発表されるが、11人の署名者はほとんどがノーベル賞受賞者である。湯川はこれに署名している。欧米人以外の署名者は湯川のみである。その後科学者の原水爆禁止運動は発展する。1955年8月6日に広島で第1回原水爆禁止世界大会が開催され、原水爆禁止日本協議会（原水協）が1955年9月発足した。日本被団協も1956年8月に長崎で発足する。湯川は時代の大局を見通せた。その目は、核力の真の原因を見通した目である。1957年第1回パグウォッシュ会議出席、1962年第1回科学者京都会議出席・・・と活動は国外・国内においてさらに広がっていく。

## 第9章　素領域理論と科学者の平和運動と世界連邦構想

　湯川の核兵器とのかかわりの源流をさぐると戦時中にさかのぼる。湯川は弟子で共著論文などで中間子論構築に協力した小林稔やアジアで最初の加速器を建設した原子核物理学の荒勝文策らとともに、戦時中、海軍の依頼で原爆にまつわる研究に関与している。しかし、最近出版された日本の荒勝文策を中心とする原子核物理学研究の黎明期の丹念な資料にもとづく歴史的分析[99]

によると、素粒子の基礎研究に関心が強かった湯川は「原爆の研究には積極的ではなかった」[99]とある。京都大学の物理学研究室での原爆開発に関連する核研究に関わる若い研究者が描かれている日米合作映画『太陽の子』(NHK 2020年8月15日放映)が話題をよんだが、荒勝研究室をモデルにしている。東京では、朝永振一郎はレーダー研究などの軍事研究に関与し、2008年ノーベル物理学賞を受賞した米国籍日本人の南部陽一郎(1921-2015)も大阪府・宝塚で陸軍の軍事研究に参画したと語っている[123]。

湯川が敗戦後初めて公にむけて書いた論考は「週刊朝日」(1945年11月)掲載の「静かに思う」であり、そのなかで「器械を制御すべき人間が、かえって器械に圧倒されはせぬかという危惧の念さえ抱かされる」[124]と述べている。「原子爆弾」という言葉は2回でてくるが、直接に核廃絶、平和運動に言及する発言はない。「器械に圧倒されはせぬか」の「器械」には暗に原子爆弾が含まれているように思われる。この論考が1年後に本『自然と理性』[124]に採録されるに当たって、文末に「付記」が書かれている。「終戦後二カ月ほどの間、いろいろな新聞や雑誌からの原稿の依頼を固くお断りして沈思と反省の日々を送って来た。その間に少し気分が落ち着いて来たので筆を執ったのがこの一篇である。一年後の今日から見るとまだまだ反省が足りないが、その時の気持ちがある程度まで現れているので採録することにした。(一九四六年十一月)」[124]とある。

1948年4月19日「新大阪新聞」に書かれた「二十世紀の不安」[125]と題する論考は原子力の不安と平和利用について触れられ、「自然のなかに潜む最も大きな力である所の原子力を平和的目的に活用するために全面的に協力することによって、初めて二十世紀の不安が除かれ、私どもの世紀が絶望の世紀から希望の世紀に転換されることを期待できるのである」と締めくくっている。

湯川秀樹著作集(全10巻・別巻1)の第5巻は『平和への希求』[126]である。科学者の著作集・全集の中でもこのように平和運動に関わるものがある人はあまり知らない。科学者の著作集では寺田寅彦の全集がよく知られているが、平和運動を一つの巻として取り上げたものはない。朝永振一郎は湯川とともに核廃絶運動に加わり、著作集に『科学者の社会的責任』[127]がある。湯川の弟子の坂田昌一(1911-1970)[128][129]、武谷三男 [8][130]も核廃絶運動に湯川・朝永とともに参加し著作もある。湯川の核廃絶・平和運動を支えた理論家である豊田利幸(1920-2009)[131]、牧二郎(1929-2005)[132]、田中正(1928-2019)[133]など、みな素粒子・原子核の理論物理学者である。学生時代に湯川に憧れて京都大学に進み、原水爆反対の運動の列に加わり、

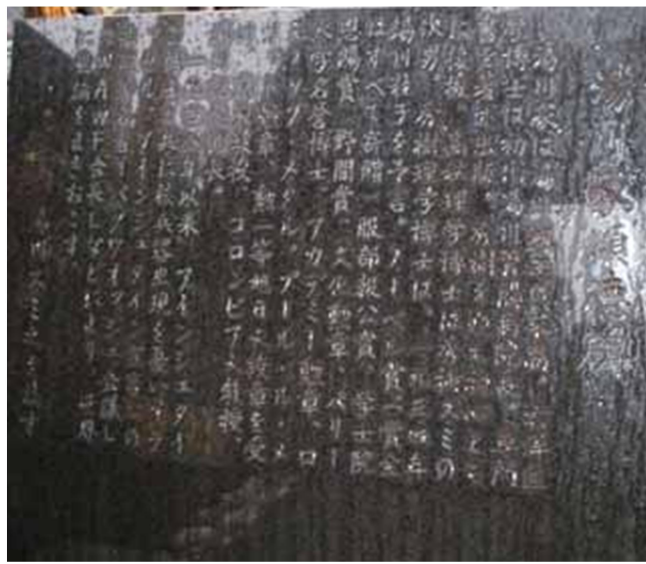

図21 「湯川家」墓所(京都東山)にある湯川スミによる湯川頌徳碑(筆者撮影)

湯川の影響を受けた学生もたくさんいる。京都大学の実験物性物理学者の加藤利三(1932－)[134]や小林稔研究室の原子核理論物理学者の永田忍(1928-2003)などもこのなかにある。こうして、次世代の若い研究者にも影響を与え、核兵器廃絶の運動は受けつがれていった。湯川の思想と運動は1954年3月の毎日新聞寄稿文以来、理論的にも実践的にも核廃絶・平和運動の先頭を切る位置にあった。核力の中間子論の発表で、世界的なボス的物理学者であるボーアやパウリからの批判にもかかわらず、極東という「辺境の地」から素粒子物理学を創業し、核力研究・原子核研究で世界をリードしてきた湯川の一貫した毅然とした「湯川精神」[135][136]の姿と重なる。

海外で核廃絶・平和運動に積極的に参加した科学者では1954年ノーベル化学賞受賞のポーリング(L. Pauling、1901-1994)が知られている。彼は原爆開発のマンハッタン計画への参加を計画責任者のオッペンハイマーに要請されながらも、平和主義者であることを理由に拒否した。核実験反対の運動で1962年ノーベル平和賞を受賞している。

科学者による核廃絶・平和運動が起こるのは第2次大戦後である。第1次大戦ではヨーロッパの各国、イギリス、フランス、ドイツは科学者も巻き込んで自国を擁護し互いに非難しあっていた[137]。量子論建設、原子物理学で指導的立場にありゾンマーフェルト学派を形成したドイツの物理学者ゾンマーフェルト(A. J. Sommerfeld、1868-1951)の第一次大戦における戦争協力・推進の言動は、文献[138]にも紹介されている。第2次世界大戦においても、イギリス、ドイツ、アメリカなど各国で科学者が軍事研究に動員され、原爆開発はその極みとも言えよう。1945年8月6日広島におけるウラニウム原爆リトル・ボーイの投下と8月9日の長崎へのプルトニウム原爆ファットマンの投下は、素粒子・原子核・物性物理学など基礎科学の物理学者が戦争協力で決定的な役割を担うことを示した。バナール(J. D. Bernal、1901-1971)は『歴史における科学』[139]において「戦争と科学」の項目をもうけ科学と軍事研究を論じている。

2度の原爆投下後も日本では本格的な原爆反対の国民的運動は起こらなかった。敗戦で明日をも知れない生活で、精神的にも肉体的にも打ちひしがれ、しかも国の独立・主権を奪われ他国の支配をうけるという日本の歴史上経験したことのない屈辱のなかで核兵器反対の思いを運動に起こすどころではなかっただろう。1949年11月3日に発表された湯川秀樹のノーベル賞受賞は打ちひしがれていた国民に大きな勇気と希望と自信を与えた。湯川は国民の希望の星であった。湯川は戦前の中間子論の研究が認められ、戦後まもなくの1948年9月から原爆開発の責任者オッペンハイマーに招かれてアメリカに渡り、プリンストン高等研究所に行った。日本は原爆を落とされ独立を失いアメリカの占領支配下にあったが、敗戦国とはいえ湯川は卑下することは何もなかった。ノーベル賞をとり、自立した国にあることを示した。湯川はアメリカから学ぶのではなくアメリカに教えるために招かれたのである。明治時代に科学の後進国として「日本人には科学は出来ない」とドイツ人・ベルツ(E. von Bälz、1849－1913)に宣告・揶揄され無念の思いで『妄想』[140]を書いた森鴎外(1862-1922)の思いが想起される[141]。1年間のプリンストン高等研究所滞在の後は、ニューヨークのコロンビア大学に移り教鞭を執った。湯川はアメリカ滞在中オッペンハイマーをはじめ誰とも原爆やそれに関わる政治の話題は意図的に避け一切話さなかったとのちに述べている[89]。湯川のこの決意は非常に固いものであった。日本が他国の支配を終え独

立し主権を回復するのは、1951 年 9 月吉田茂 (1878-1967) 全権代表が連合国と締結したサンフランシスコ平和条約が 1952 年 4 月 28 日に発効してからである。これにより原爆投下による敗戦、他国による占領支配、食糧難と極度の困窮という屈辱的なみじめな状況を脱する道へ踏み出すことができた。湯川の次男でコロンビア大学で物理学を修めた物理学者の湯川高秋 (1934-1971) は、日本に帰り講談社で吉田茂の平和への思いを世界に伝えたいとその著書の翻訳に携わるが、父親・秀樹の平和の希求に影響をうけたであろうか、その活動は湯川の思いを考えると理解できるような気がする。彼は若くして急死するが、その前日まで翻訳に携わっていたという [24]。湯川が 63 歳で 1970 年に京都大学を定年退職した翌年、次男の突然の死である。湯川自身も死の直前まで核廃絶・平和の希求に自らの命を減らす覚悟で献身した。

　湯川は核廃絶への取り組みと同時に自らの非局所場理論を発展させ、1968 年ころまでに素領域理論を定式化する。湯川の学問への情熱は決して衰えなかった。湯川は中間子の存在を予言し素粒子物理学を創業したが、湯川の予言した核力を引き起こす π 中間子の他にもたくさんの中間子が見つかった。その π 中間子の質量は、その後発見されたたくさんの中間子の中でも最も質量が小さい。米シカゴ大学の南部陽一郎は 1957 年にバーディン (J. Bardeen、1908-1991)、クーパー (L. Cooper、930-2024))、シュリーファー (J. R. Schrieffer、1931-2019) により発見された超伝導理論 (BCS 理論) を素粒子に適用し、1960 年、素粒子における対称性の自発的破れを提唱し、陽子・中性子などの質量獲得機構や π 中間子の質量が小さい理由を説明した。中間子の中には荷電パリティ (CP) 保存則を破るものがあったが、京都大学にいた益川敏英 (1940-2021) と小林誠 (1944-) は牧二郎らの素粒子の 4 元模型を発展させ、1973 年に 6 元模型で CP 対称性の破れを説明した。湯川の中間子による核力、強い相互作用の解明は、重力を発見した 英国のニュートン (I. Newton、1643-1727)、電磁気力の理論を完成した英国のマックスウェル (J. C. Maxwell、1831-1879)、弱い相互作用の理論のイタリアのフェルミ (E. Fermi、1901-1954) と並び、自然界で知られている 4 つの相互作用の 1 つを発見した歴史的業績である。湯川以外はすべて近代科学が勃興し発展してきたヨーロッパ人である。湯川は自らが開いた素粒子物理学の個別的な課題よりも、より根源的な課題に向かった。素粒子物理学は中間子や陽子・中性子がさらに小さい物からできていると考え、それらの探索と、それらの物質の性質やそれらに働く相互作用と法則の探求にむけて、要素還元主義的に進んできた。ギリシアの哲学者デモクリトス (Democritus、B.C.460 頃-B.C.360 頃) が物質の最小単位として考えた「アトム」探求の道である。

　だが湯川は違った。デモクリトスは空虚のなかにアトムがあるとした。しかし空虚、今日の物理学の言葉では「真空」、言い換えると何もない「空間」の何たるかは全然わかってない。アインシュタインは相対論で時間 (time) と空間 (space) を同等にあつかい「時空」(spacetime) の概念を生み出した。湯川は天体、分子、原子、素粒子などの宇宙の各階層の物質の性質や法則、それらの従う運動法則だけでなく、より根源的問題である、それら「万物」が共通に棲む「入れ物」である「時空」の何たるかを知りたかった。そして、定式された理論が素領域理論である [142]。湯川のこの考えを表現した有名な和歌がある。「天地(あめつち)は逆旅(げきりょ)なるかも鳥も人もいづこよりか来ていづこにか去る」[143] (巻末最後から 2 番目の 1 首)。筆者が 1969 年京都大学卒業時に湯川先生

よりいただいた色紙には「天地萬物逆旅　光陰百代過客　湯川秀樹」とある。拙稿[1]で紹介したように、湯川が好んだ中国の詩人李白(701-762)の「春夜宴桃李園一序」にある「夫天地者萬物之逆旅　光陰者百代之過客」による。湯川の素領域理論の世界観をよく表している。湯川は4次元ミンコウスキー時空で非局所場統一理論をつくろうとした。アインシュタインが晩年、重力と電磁気力の統一をめざして統一場理論の建設にのめり込んだ姿勢と重なる所があるように筆者には思える。アインシュタインの統一場理論は、1970年代初頭に湯川の核力がゲージ相互作用（量子色力学）であることが確立されて後、現実的課題となるのである。ランドール (L. Randall、1962-) とサンドラム (R. Sundrum、1964-) [144]は4つの力（重力・電磁気力・強い力・弱い力）の階層性問題を高次元時空で統一できるとし、時空はワープした高次元（5次元）反ドジッター時空であった。21世紀になってからの最近の重力と量子場のホログラフィーによる研究の進展で「時空」の理解が現実的課題となってきている[145]。湯川は半世紀以上も前にその課題に挑戦し続けていたのである。まさに天才の取り組む「課題意識」である。時代を一回りも二回りも先取りした挑戦であり、アインシュタインの統一場理論とともに成功するには早すぎた。湯川の前の世代で、原子物理に挑戦して、英国のブラッグ父子とともにX線の結晶回折を世界に先がけて示し、日本のX線結晶物理学を創出した寺田寅彦の弟子である物理学者の渡辺慧(1910-1993)は、湯川追悼特集記事[146]で湯川についてこう述べている。少し長くなるが本質的なことを言っているので、引用しよう。「これは（素領域理論）は、成功した理論ではありません。＜略＞科学史家は普通成功した論文、オーソドックス化した理論だけ書きますが、私はそういう論文よりももっと大きな貢献をするのは、失敗した論文、外典(apocrypha)化した理論、中絶した試みだと思います。・・・その失敗した理論に本質的に正しいものが含まれていて、失敗はその本質とは無関係な他の要素に帰着することがしばしばあるからです。・・・そういう論文にはしばしばなにか有益な"哲学"ないしは"示唆"が含まれている場合が多いのです。湯川さんの非局所場は、明確な"哲学"をもっており、数学的にも独創的であり、その真実性は、実験データに直接合わなくても、人を打つものを持っています。今は認められていなくても、やがて別の衣を着せられて甦って来るでしょう。これは中間子論以上の大事業です。こういう独創的な仕事はノーベル賞にはなりません」。21世紀になり、いまや素粒子物理学、宇宙理論では「時空」論が大流行である。物理学に限らず、数学の分野でも数字が0と1の2個しかないブール代数をつくった英国人ブール (G. Boole、1815-1864) を思い起こさせる。彼が研究したのは19世紀前半である、その数学が広く認められ不可欠な数学として威力を発揮するのはコンピュータが登場する20世紀中ごろになってからである。100年も先行していた研究である。

　筆者は1967-1968年湯川の「物理学通論」の講義を京都大学で聞いた。湯川は当時出て流行っていた米国のゲルマン (M. Gell-Mann、1919-2019) のクォーク模型には全く興味を示さなかった。講義でもほとんど触れなく、クォーク模型を批判していた。（講義の一コマは来日中のハイゼンベグ (W. Heisenberg、1901-1976) による特別講義「物質講義についての哲学」であったが、彼も思考が哲学的でクォーク模型には触れなかった）。ゲルマン自身も分数電荷のクォークは数学的な記号で物理的実体のないものだと云っていた。ノーベル賞を受賞した科学者の多くは個別科

学で大きな進展を達成した。シュレーディンガー (E. Schrodinger、1887-1961) の波動方程式、ハイゼンベルグの行列力学、ボーアの原子模型も偉大な研究成果だ。ミクロの原子階層での物質の従う法則を求めたものである。物質と時空を包括し統一することを目指すものではない。ゲルマンのクォーク模型やクォーク間力の漸近的自由性、超伝導理論も特筆されるべき偉大な研究成果だが、それは物質や力の性質の理解である。物質と時間と空間の統一的理解を目指した研究ではない。

湯川は中間子論で原子核が存在する基本的原理を明らかにしたが、その後の研究では原子核の個別的研究には携わらなかったし、素粒子の個別的研究にも携わらなかった。ゲルマンのクォーク模型などには全く興味を示さなかった。ニュージーランド出身の英国人ラザフォードはギリシア人デモクリトスが考えたアトムを破壊し、アトムが内部構造を持つことを実験的に示した偉大な科学者である。英国のディラック (P. Dirac、1902-1984) はアインシュタインの相対論の式を"因数分解"し相対的量子力学をつくり反粒子の存在を示した。英国人のマックスウェルは電気と磁気の統一理論を作り電磁波を予言し、光がその一種であることを示した。湯川は物質と時空の統一理論と同時に人間社会の平和な統一理論を目指した。

湯川は京都東山に眠る。湯川スミ夫人が記した「湯川家頌徳碑」(図 21) では、湯川秀樹の功績を讃えている。「一九四八年以来アインシュタイン博士と共に核兵器出現を憂い『ラッセル・アインシュタイン宣言』の共同署名者『パグウォッシュ会議』『WAWF 会長』などにより、世界に世論をまきおこす。湯川スミ之を記す」とある。湯川の核廃絶運動を支えてきたスミ夫人の思いが込められている。

湯川の平和への活動が 1948 年以前からであると記述していることに注目したい。一般には湯川が核廃絶の社会的発言をするのは、1954 年 3 月 30 日毎日新聞の寄稿が最初とみなされている。しかし、スミ夫人の記述は高知訪問中も、それ以前も、湯川本人が核兵器廃絶の意識を持っていたということを強調している。ビキニ水爆に言及し、しかも公然と反対することは、日本を占領支配したアメリカに公然と反対の意思を表示することである。「天皇について有名」[11] とパイスが記したように日本を代表するノーベル受賞物理学者湯川にとって非常に勇気のいることであった。

「洞窟の蝙蝠」は世界に飛び立つこととなった。鍾乳洞の蝙蝠を湯川が見たのは、夜須小学校での除幕式、城山高校での 2 度の講演を終えた後だった。歌人でもある湯川はその蝙蝠に自分を重ねた。2 日目の午前は親友と桂浜の坂本龍馬の大きな銅像を見上げたであろう。そして洋々たる太平洋を夫人とともに眺めた。龍河洞の蝙蝠と自己を重ねた湯川は龍馬の像と自己をどう重ねたであろう。龍馬は一介の郷士でありながらも、中浜万次郎から聞き取りをした河田小龍からアメリカの文明・世界の広さを聞き、討幕に奔走し明治維新の扉を開こうとした。中浜万次郎は土佐沖を流れる黒潮に流され、一介の漁師からアメリカ文明を学び、龍馬に託した。湯川は荘子とともに空海 (774-835) を好んだ [1] [147] が、その空海は四国 88 か所の巡礼の道を開いた。太平洋の荒波をうけ気象の厳しい土佐は「修行の道場」と言われる。湯川は龍河洞を見、桂浜の太平洋と龍馬を見たあと、児童に「洞窟の蝙蝠」の話をした。湯川は児童に向かって話したが、湯川を

注視している報道陣がいることはよく知っている。湯川は自己を蝙蝠に暗喩し世に向けて話したのである。いや、むしろ自らに向けて話したのかもしれない。湯川は自分自身を子どもたちにかさね、自らを諭すかのように子どもたちに話かけたようにも思える。高知で最後に湯川は念願の闘犬も見た。化粧まわしをした土佐犬は優雅で力強い。湯川は自らを土佐犬にも重ねただろうか。龍河洞、桂浜、土佐犬、土佐の友人、人々・・・、土佐の気風と風土が湯川の背をおしたことは確かなようだ。

日本で平和運動にかかわった科学者は沢山いる。だが、ノーベル賞受賞者であったとしても湯川のように夫人も加わって核廃絶・平和の希求のために生涯を捧げた科学者はほかにいるだろうか。湯川が死の直前まで、核廃絶のための活動し、病気をおして車椅子に乗って、科学者の平和運動に身を挺した姿は忘れられない。高知に生まれ育った筆者は時代と運命のめぐりあわせか湯川に憧れて京都大学に進み、湯川の講義を京都大学できき、大学院で研究者としてその研究に接し、その深遠な学問とひたむきな探求の姿勢に感銘を受けた。湯川の終生かわらぬ学問への探求の精神と献身的な平和の探求の姿勢は忘れられない。

## 第 10 章　終章

湯川秀樹先生が学問一筋の学究生活から偶然にも高知訪問直前にアメリカのビキニ水爆実験が行われ日本人の被害者が出るという歴史的な大事件に遭遇し、苦悩の中でいかに核廃絶と平和の希求に後半生をかける決断を一晩という短期間にすることができたのか、その背景をくわしく考察した。その背後には、高知訪問が湯川の決断を後押しするいくつかの要因があることを明らかにした。湯川の友人、土佐の歴史と風土があった。その後の活動に湯川夫人のサポートもあった。しかし、何よりも大局的な見通し、勇敢な決断ができる湯川の人格・思想・生き方そのものがあることを論じた。

高知訪問で湯川は夜須小学校の胸像除幕式のあいさつで「役にたてばうれしい」と語った。素粒子物理学を切り開いた湯川だが、その学問は浮世離れした学問のように考えられがちである。しかし、湯川の根底には「役に立ちたい」という願いがあった。湯川胸像建設から 70 年、2024 年 12 月 2 日高知県香南市夜須町の夜須中学校で、70 周年を記念して「第 1 回湯川胸像建設 70 周年夜須科学セミナー」が夜須中学・小学校 5・6 年生の授業の一環として行われた[148]。日本ではじめて建てられ、湯川夫妻が出席し除幕された湯川秀樹胸像は科学探究と平和の希求の像としていまある。湯川の決断した核廃絶と平和の願いも地域住民・子どもへと受け継がれている。

2024 年ノーベル平和賞は原水爆反対・核廃絶の活動を牽引し世界から一切の核兵器の廃絶の運動を行ってきた、湯川の「高知での決意」から 2 年後 1956 年 8 月に設立された日本原水爆被害者団体協議会（日本被団協）に贈られた。湯川に 1949 年原子核物理学・中間子論の研究でノーベル物理学賞を贈ったノーベル財団は 75 年を経て日本の核兵器廃絶運動にもノーベル平和賞を贈ったのである。湯川が夫人とともに後半生を捧げた核兵器廃絶運動である。湯川は泉下の人となり京都東山・知恩院にスミ夫人とともに眠る。だが、突然遭遇したビキニ水爆実験・第五福竜丸被爆事件で湯川が苦悩し、核廃絶へと一夜のうちに「生涯の決断」をした高知に、湯川胸像は科学と

平和の不滅の象徴のごとく今もたつ。

## 謝辞

## Summary

In 1954, following a five-year research period in the U.S., Professor Hideki Yukawa returned to Japan and visited Kochi on March 21 to attend the unveiling ceremony for the first statue of him ever built in Japan, a project initiated by the PTA of Yasu Elementary School in Yasu Town, Kochi Prefecture. By a coincidence of history, just three weeks prior on March 1, the U.S. had conducted a hydrogen bomb test at Bikini Atoll in the Pacific Ocean. Many Japanese fishing boats were operating there at the time and had not been informed in advance. As a result, numerous boats, including the Daigo Fukuryu Maru, were exposed to radiation. Upon his arrival at Kochi Station on the evening of March 21, Yukawa was relentlessly questioned by reporters about the Bikini hydrogen bomb. This was a source of deep anguish for Yukawa, a Japanese physicist who had won the Nobel Prize for his work on "atomic physics." He firmly refused to answer, stating that the topic was "outside the scope of my research." The next evening, at a public lecture in Kochi City on March 22, he again refused to speak about the Bikini hydrogen bomb or nuclear power, stating that he was an amateur in nuclear research and that there were many other experts. However, just four days later, on March 28, after returning to Kyoto, Yukawa drafted his famous essay, "The Turning Point for Humanity and Atomic Power," which was published in a newspaper on March 30. From that point on, he was drawn into the tumultuous issue of the Bikini hydrogen bomb and nuclear power. When did a tormented Yukawa make his decision? This article meticulously reveals, based on historical documents, what led the anguished Yukawa to make such a rapid decision within a single day and what caused the immense change in his mindset overnight.

（2025 年 9 月 30 日）

## 参考文献

ヨドバシカメラの位置。現存せず）で『綜合原爆展』を 10 日間の会期で開催した」。**加藤**の証言：「1954 年 3 月 1 日にアメリカが太平洋上のビキニ環礁で水爆実験をして、160 km 離れたところで操業していた焼津のマグロ漁船第五福竜丸が大量の『死の灰』を浴び、乗組員 23 名全員が被ばくしました。＜略＞理学部の荒勝研究室は、広島への原爆投下後、すぐに医学部の研究者と一緒に現地調査に行った経験をもっていました。その頃、同じクラスの永田忍さんらが水爆展をしようと提案し、クラスの有志で取り組むことになりました。＜略＞夜中にみんなでパネルを作って、府立大学のグランド（？）で行われたメーデーの集会に合わせて、グランドの入り口付近に持ち込んで展示会をしました。私は理学部の学生自治会の委員をしていたので、西部講堂の前の広場に集まってみんなを率いてデモに行きました。その写真もあります」。